\documentclass[final]{article}%
\usepackage[margin = 1.5in]{geometry}
\usepackage{graphicx}
\usepackage{natbib} 
\usepackage{amsmath}
\usepackage{amsfonts} 
\usepackage{mathtools}
\usepackage{bm}     	
\usepackage{mathrsfs}  

\usepackage{xcolor}   
\usepackage{hyperref} 
\usepackage{undertilde} 
\usepackage{booktabs} 
\usepackage{multirow} 
\DeclarePairedDelimiter\abs{\lvert}{\rvert}%
\DeclarePairedDelimiter\norm{\lVert}{\rVert}%
\newcommand{\Sum}[2]{\displaystyle\sum^{#1}_{#2}} 

\newcommand{\RN}[1]{%
  \textup{\uppercase\expandafter{\romannumeral#1}}%
}				
\newcommand{\dsim}{\overset{\mathcal{D}}{\sim}}

\usepackage{lipsum}

\newtheorem{theorem}{Theorem}

\newtheorem{corollary}[theorem]{Corollary}

\newenvironment{proof}[1][Proof]{\textbf{#1.} }{\ \rule{0.5em}{0.5em}}

\usepackage{bm}
\usepackage{sectsty}

\DeclareMathOperator*{\argmin}{arg\,min}

\usepackage{multirow}

\usepackage[title]{appendix}
  
\begin{document}

\title{\textsc{Adaptive Random Bandwidth for Inference in CAViaR Models }}
\author{\textbf{Alain Hecq}
\and \textbf{Li Sun\thanks{Corresponding author: Li Sun, Maastricht University, School of Business and Economics,
Department of Quantitative Economics, P.O.Box 616,
6200 MD Maastricht, The Netherlands. Email: l.sun@maastrichtuniversity.nl}} }
\date{Maastricht University \\[0.38cm] January 29, 2021}
\maketitle

\begin{abstract}
\noindent This paper investigates the size performance of Wald tests for CAViaR models~\citep{engle2004caviar}. We find that the usual estimation strategy on test statistics yields inaccuracies. Indeed, we show that existing density estimation methods cannot adapt to the time-variation in the conditional probability densities of CAViaR models. Consequently, we develop a method called \textit{adaptive random bandwidth} which can approximate time-varying conditional probability densities robustly for inference testing on CAViaR models based on the asymptotic normality of the model parameter estimator. This proposed method also avoids the problem of choosing an optimal bandwidth in estimating probability densities, and can be extended to multivariate quantile regressions straightforward.
\bigskip



\end{abstract}

\section{Introduction}
Financial risk management is at the heart of banks' and financial
institutions' activities to guide them in their investment plans, supervisory
decisions, risk capital allocations and for external regulations. The use of quantitative risk measures has become essential in financial risk management. One of the most popular risk measures associated with financial portfolios is the value at risk (VaR hereafter). The VaR at probability $\tau \in (0,1)$ of a portfolio is defined as the minimum potential loss that the portfolio may suffer in the worst $\tau$ portion of all possible outcomes over a given time horizon. VaR is very intuitive~\citep{duffie1997overview} and has for instance been incorporated into \textit{the 1996 Amendment to
the Capital Accord} for measuring the market risk in financial positions of each financial institution. Therefore, VaR is still a
widely used risk measure even though many approaches to
measuring market and credit risks have been proposed in the
literature.

Generally, there are three ways to estimate VaR: \textit{(i) }%
historical simulations, \textit{(ii)} semi-parametric approaches and \textit{%
(iii)} fully parametric frameworks. Within the class of semi-parametric
approaches, it typically includes extreme value theory analyses and quantile regression techniques. In this paper, we focus on quantile regressions for the VaR estimation as quantile regressions are straightforward in studying one quantile of interest and numerically efficient without imposing parametric distributional assumptions.

Despite that the VaR is just a particular quantile of future portfolio losses conditional on present information, it is essentially a part of the underlying conditional distribution. VaR models are supposed to embrace features of the empirical conditional distributions of returns, such as time-variation and conditional heteroskedasticity. Drawing on (G)ARCH specifications which capture the presence of time-varying conditional heteroskedasticity in time series, \cite{engle2004caviar} have proposed to estimate conditional autoregressive value at risk by regression quantiles (CAViaR). It is appealing to consider CAViaR models for estimating VaR as CAViaR models associate the conditional quantile of interest with observable variables as well as the implicit information on lagged conditional quantiles.

This paper carefully investigates the size performance of Wald tests for CAViaR models. Having an accurate test statistic is important to obtain reliable models in financial applications. Several specifications are nested within a CAViaR specification, such as static quantile regressive models and quantile autoregressive models \citep[see][]{koenker2006quantile,hecq2020selecting}. Moreover, there exists several models nested within the general CAViaR specification that have been proposed in the literature. For instance, asymmetric slope CAViaR models~\citep{engle2004caviar} that split the effect of positive and negative yesterday's news shocks. Wald tests are used to test the null of a symmetric news impact. However, we find that the usual estimation strategy yields inaccuracies. Indeed, we show that existing density estimation methods cannot adapt to the time-variation in the conditional probability densities of CAViaR models. The method that we develop in this paper is able to adapt to time-varying conditional probability densities and produces much more reliable results than the existing ones for inference testing on CAViaR models based on the asymptotic normality of the model parameter estimator. This proposed method also avoids the haunting problem of choosing an optimal bandwidth in estimating probability densities, and can be extended to multivariate quantile regressions straightforward in theory.

The remainder of this paper is structured as follows.
In Section~\ref{sec:intro_CAViaR}, stability conditions for CAViaR data generating processes (DGPs) to be non-explosive are derived. In
Section~\ref{sec:covariance_mat_estimation}, we investigate the size performance of Wald tests for CAViaR models and find large size distortions by the usual estimation strategy. So we introduce a method called \textit{adaptive random bandwidth}. An empirical study on stock returns is
performed in Section~\ref{sec:empirical_study}. Finally Section~\ref%
{sec:conclusion} concludes this paper.

\section{The CAViaR model}

\label{sec:intro_CAViaR} Let us consider a stationary time series process $%
\left\{ y_{t}\right\} _{t=1}^{T}$ for instance the return of an asset
or a portfolio, and denote $\boldsymbol{x}_{t}$ a vector of observable
variables at time $t$ and $\mathcal{F}_t$ the information set up to time $t$ which is the $\sigma$-algebra generated by $\left\{\boldsymbol{x}_{t}, y_t, \boldsymbol{x}_{t-1}, y_{t-1}, \ldots \right\}$. The $\tau$-th quantile ($\tau \in (0,1)$) or the opposite $\text{VaR}_{\tau}$ of $y_{t}$ conditional on $\mathcal{F}_{t-1}$ is denoted as $
f_{t}(\boldsymbol{\beta }_{\tau },\boldsymbol{x}_{t-1})$ (or simply $f_{t}(\boldsymbol{\beta _{\tau }})$ when $\boldsymbol{x}_{t-1}$ is taken in obviously). A generic CAViaR
specification proposed by \cite{engle2004caviar} is 
\begin{equation}
f_{t}(\boldsymbol{\beta }_{\tau })=\beta _{0}+\sum\limits_{i=1}^{q}\beta
_{i}f_{t-i}(\boldsymbol{\beta }_{\tau })+\sum\limits_{j=1}^{r}\beta
_{q+j}\,l(\boldsymbol{x}_{t-j}),
\label{eq:CAViaR}
\end{equation}%
where $\boldsymbol{\beta _{\tau }}^{\prime }:=\left[ \beta _{0},\beta
_{1},\ldots ,\beta _{p}\right] $ collects the $p=q+r$ slope parameters, and $l$ is
a function of a finite number of lagged observable variables, for instance
the lagged returns entering potentially with different weights for positive
and negative past lagged returns. As described in \cite{engle2004caviar} the
autoregressive terms $\beta _{i}f_{t-i}(\boldsymbol{\beta }_{\tau })$
can ensure that the quantile changes smoothly over time. The quantile
autoregressive model (QAR) of Koenker and Xiao (2006) is nested in the
CAViaR specification by restricting $\beta _{1}=...=\beta _{q}=0$ in CAViaR. The
role of $l(\boldsymbol{x}_{t-j})$ is to account for the association of $f_{t}(\boldsymbol{\beta }%
_{\tau })$ with observable variables in $\mathcal{F}_{t-1}$. CAViaR models as a generalization of QAR models are able to capture the time-variation in the conditional quantile in a way similar to GARCH models in explaining time-varying volatility and volatility clustering in financial time series in addition to ARCH models.

The CAViaR model~\eqref{eq:CAViaR} is nonlinear in parameters as long as there exists a nonzero
$\beta _{i},i\in \left\{ 1,\ldots ,q\right\} $ which leads to $\frac{\partial
f_{t}(\boldsymbol{\beta }_{\tau })}{\partial \beta _{i}}=f_{t-i}(\boldsymbol{%
\beta }_{\tau })+\beta _{i}\frac{\partial f_{t-i}(\boldsymbol{\beta }_{\tau
})}{\partial \beta _{i}}$ not independent of $\beta _{i}$.\footnote{In Appendix~\ref{sec:nonlinearity_CAViaR}, the gradient and the Hessian matrix of CAViaR models are illustrated to emphasize that the nonlinearity of model parameters makes CAViaR models different from other linear quantile regression models.} The algorithm to estimate CAViaR models is given in Section~\ref{sec:CAViaR_Algorithm}.

\begin{figure}[hbtp]
\centering
\hspace*{-1cm}\includegraphics[width = 1.1\textwidth, height = 7cm]{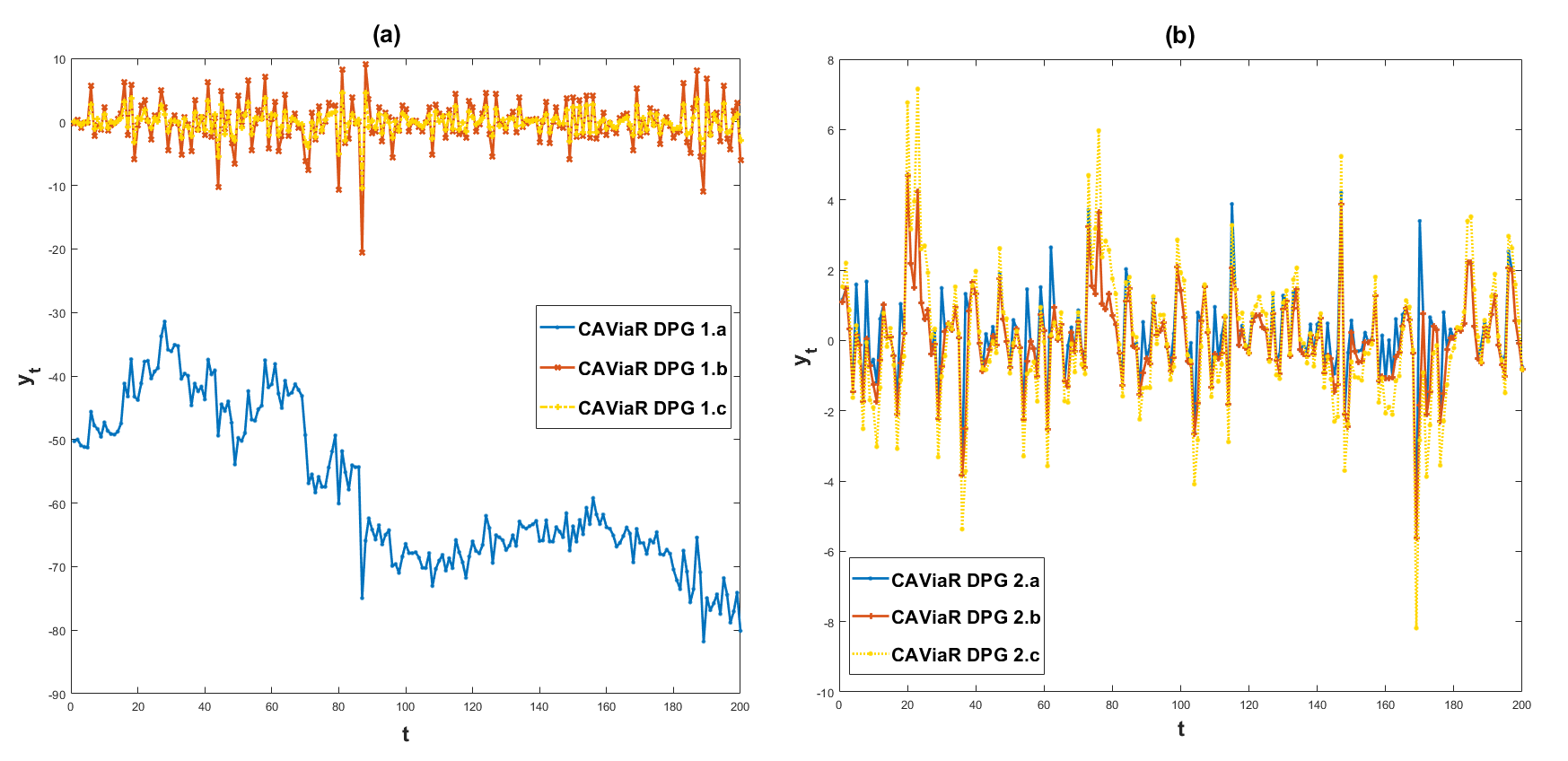}
\caption{Time series plots of CAViaR DGP samples}
\label{fig:ts_simulations_1}
\end{figure}

For illustration, we simulate samples from the following three CAViaR DGPs in~\eqref{eq:CAViaR_samples_Fig1_DGPs} and plot Figure~\ref{fig:ts_simulations_1} (a). \footnote{All the simulations of CAViaR DGPs in this paper follow the procedure given in Appendix~\ref{sec:simulate_CAViaR_DGPs}.} In Figure~\ref{fig:ts_simulations_1} (a), we see a decreasing trend in CAViaR DGP 1.a mainly due to the negative term $-0.5|y_{t-1}|$ in $f_{t}(\boldsymbol{\beta }_{\tau})$ compared with CAViaR DGP 1.b. Comparing CAViaR DGP 1.b with 1.c, we find that CAViaR DGP 1.b has a larger spread due to a higher slope of $ f_{t-1}(\tau)$ in $f_{t}(\boldsymbol{\beta }_{\tau})$. A similar finding further applies on Figure~\ref{fig:ts_simulations_1} (b) which plots simulated samples of CAViaR DGP 2.a, 2.b and 2.c in~\eqref{eq:CAViaR_samples_Fig2_DGPs} respectively.
\begin{equation}
\left\{
\begin{aligned}
\text{CAViaR DGP 1.a:}\qquad 
f_{t}(\boldsymbol{\beta}_{u_t} )
				& = F_{t(3)}^{-1}(u_t)+ 0.5\,f_{t-1}(\boldsymbol{\beta }_{u_t})-0.5|y_{t-1}|,
				\\
\text{CAViaR DGP 1.b:}\qquad 
f_{t}(\boldsymbol{\beta}_{u_t} )
				& = F_{t(3)}^{-1}(u_t)+ 0.5\,f_{t-1}(\boldsymbol{\beta }_{u_t})-0.5\,y_{t-1},
				\\
\text{CAViaR DGP 1.c:}\qquad
f_{t}(\boldsymbol{\beta}_{u_t} )
				& = F_{t(3)}^{-1}(u_t)-0.5\,y_{t-1},				
\end{aligned}
\right.
\label{eq:CAViaR_samples_Fig1_DGPs}
\end{equation}
where $\{u_{t}\} $ is i.i.d. in the standard uniform distribution (denoted as $\mathcal{U}(0,1)$) and $F_{t(3)}^{-1}(\cdot)$ is the inverse function of Student's t-distribution with $3$ degrees of freedom ($t(3)$ hereafter).

\begin{equation}
\left\{
\begin{aligned}
\text{CAViaR DGP 2.a:}\qquad 
f_{t}(\boldsymbol{\beta}_{u_{t}} )
				& = F_{t(3)}^{-1}(u_{t}) - 0.5\,f_{t-1}(\boldsymbol{\beta }_{u_{t}})+0.5|y_{t-1}|
				\\
\text{CAViaR DGP 2.b:}\qquad 
f_{t}(\boldsymbol{\beta}_{u_{t}} )
				& = F_{t(3)}^{-1}(\tau) - 0.5\,f_{t-1}(\boldsymbol{\beta }_{u_{t}})+0.5\,y_{t-1}
				\\
\text{CAViaR DGP 2.c:}\qquad
f_{t}(\boldsymbol{\beta}_{u_{t}} )
				& = F_{t(3)}^{-1}(u_{t})+0.5\,y_{t-1},			
\end{aligned}
\right.
\label{eq:CAViaR_samples_Fig2_DGPs}
\end{equation}
where $u_t \overset{i.i.d.}{\sim} \mathcal{U}(0,1)$, $t=1,2,\ldots, T$.

\subsection{The stability conditions for CAViaR models}

\label{subsec:stability condition} 
The stationarity of CAViaR time series is required for the model estimation
consistency~\citep{engle2004caviar}. After simulating a CAViaR DGP, we can view its behaviour such as explosiveness in the long run. We know that a time series is explosive if and only if at least one conditional quantile of the time series with nonzero probability density to occur is explosive. So we derive stability conditions for the conditional $\tau$-th ($\tau\in(0,1)$) quantile of a CAViaR DGP $\left\{ y_{t}\right\} $ specified as follows:
\begin{equation}
y_{t}=f_{t}(\boldsymbol{\beta }_{u_{t}})=\beta
_{0}(u_{t})+\sum\limits_{i=1}^{q}\beta _{i}(u_{t})\,f_{t-i}(\boldsymbol{%
\beta }_{u_{t}})+\sum\limits_{j=1}^{r}\beta _{q+j}(u_{t})\,y_{t-j},
\label{eq:CAViaR_AR}
\end{equation}%
where $u_t \overset{i.i.d.}{\sim} \mathcal{U}(0,1)$, and $\boldsymbol{\beta _{u_{t}}}^{\prime }:=\left[ \beta _{0}(u_{t}),\beta
_{1}(u_{t}),\ldots ,\beta _{p}(u_{t})\right] $ with $p=q+r$. There is a monotonicity requirement on this model which is that $f_{t}(%
\boldsymbol{\beta }_{u_{t}})$ is monotonically increasing in $u_{t}$ so that
the $\tau $-th quantile ($\tau \in (0,1)$) of $%
y_{t}$ conditional on $\mathcal{F}_{t-1}$ can be expressed as $f_{t}(%
\boldsymbol{\beta _{\tau }})$. 

Assume the conditional $\tau $-th quantile of $\left\{ y_{t}\right\} $
follows the model~\eqref{eq:CAViaR_AR} with nonzero probability density to occur at each time. Without loss of generality, there is a time $t\in \left\{
1,\ldots ,T\right\} $ such that 
$$
y_{t}=f_{t}(\boldsymbol{\beta }_{\tau
})=\beta _{0}+\sum\limits_{i=1}^{q}\beta _{i}f_{t-i}(\boldsymbol{\beta }%
_{\tau })+\sum\limits_{j=1}^{r}\beta _{q+j}\,y_{t-j}.
$$
Now let us derive the
value of $y_{t}$. First we have the following equation from~%
\eqref{eq:CAViaR_AR}. 
\begin{equation*}
\begin{aligned} \left( 1 - \sum\limits^r_{j=1}\beta_{q+j}\,L^{j} \right) y_t
& = \beta_0 +\sum\limits^q_{i=1}\beta_i f_{t-i}(\boldsymbol{\beta}_{\tau})
\\ & = \beta_0 + \sum\limits^q_{i =1}\beta_{i} \left( \beta_0
+\sum\limits^q_{i_1 =1}\beta_{i_1} f_{t-i-i_1}(\boldsymbol{\beta}_{\tau})
+ \sum\limits^r_{j=1}\beta_{q+j}\, y_{t-i - j} \right) \\ & = \beta_0 +
\sum\limits^q_{i =1}\beta_{i} \left( \beta_0 +\sum\limits^q_{i_1
=1}\beta_{i_1} f_{t-i-i_1}(\boldsymbol{\beta}_{\tau}) +
\sum\limits^r_{j=1}\beta_{q+j}\, L^{i + j}\,y_{t} \right) \\ & = \left( 1
+ \sum\limits^q_{i =1}\beta_{i} \right)\beta_0 + \sum\limits^q_{i
=1}\sum\limits^q_{i_1 =1} \beta_{i}\beta_{i_1}
f_{t-i-i_1}(\boldsymbol{\beta}_{\tau}) + \sum\limits^q_{i
=1}\sum\limits^r_{j=1} \beta_{i}\beta_{q+j}\, L^{i + j}\,y_{t},
\end{aligned}
\end{equation*}%
where the second line is obtained by substituting the specification~\eqref{eq:CAViaR_AR} of $f_{t-i-i_1}(\boldsymbol{\beta}_{\tau})$ into the first line, and $L$ is the lag operator. Further rewrite the above equation, and we have 
\begin{equation*}
\begin{aligned} \left( 1 - \sum\limits^r_{j=1}\beta_{j}\,L^{j} -
\sum\limits^q_{i =1}\sum\limits^r_{j=1} \beta_{i}\beta_{q+j}\, L^{i +
j} \right) y_t & = \left( 1 + \sum\limits^q_{i =1}\beta_{i}
\right)\beta_0 + \sum\limits^q_{i =1}\sum\limits^q_{i_1 =1}
\beta_{i}\beta_{i_1} f_{t-i-i_1}(\boldsymbol{\beta}_{\tau}) \\ & =
\left( 1 + \sum\limits^q_{i =1}\beta_{i} \right)\beta_0 +
\sum\limits^q_{i =1}\sum\limits^q_{i_1 =1} \beta_{i}\beta_{i_1}
f_{t-i-i_1}(\boldsymbol{\beta}_{\tau}). \end{aligned}
\end{equation*}%
We continue to rewrite the lagged terms of $f_{t}(\boldsymbol{\beta }_{\tau
})$ on the right-hand side of the above equation, and then organize the equation such that only the left-hand side contains terms of $y_t$. Therefore, we obtain that 
\begin{equation}
\resizebox{1.08\textwidth}{!}{ $ \begin{aligned} & \left( 1 -
\sum\limits^r_{j=1}\beta_{q+j}\,L^{j} - \sum\limits^q_{i
=1}\sum\limits^r_{j=1} \beta_{i}\beta_{q+j}\, L^{i + j} - \ldots -
\sum\limits^q_{i =1}\sum\limits^q_{i_1 =1}\ldots\sum\limits^q_{i_n
=1}\sum\limits^r_{j=1} \beta_{i}\beta_{i_1}\ldots\beta_{i_n}\beta_{q+j}\,
L^{i + i_1\ldots + i_n + j} \right) y_t \\ & = \left( 1 +
\sum\limits^q_{i =1}\beta_{i} + \left( \sum\limits^q_{i
=1}\beta_{i}\right)^2 + \ldots + \left( \sum\limits^q_{i
=1}\beta_{i}\right)^n\right)\beta_0 \\ & + \sum\limits^q_{i
=1}\sum\limits^q_{i_1 =1}\ldots\sum\limits^q_{i_{n+1} =1}\sum\limits^r_{j=1}
\beta_{i}\beta_{i_1}\ldots\beta_{i_{n+1}} f_{t-i-i_1-\ldots-
i_{n+1}}(\boldsymbol{\beta}_{\tau}). \end{aligned} $ }  \label{eq:poly_yt}
\end{equation}%
Now we can get the first necessary condition for $\left\{ y_{t}\right\} $ to be
nonexplosive, which is 
\begin{equation}
\left| \sum\limits_{i=1}^{q}\beta _{i} \right| <1.  \label{eq:1st_cond}
\end{equation}%
Under the condition~\eqref{eq:1st_cond}, we can simplify the equation~\eqref{eq:poly_yt} when
letting $n\rightarrow \infty $ as follows: 
\begin{equation*}
\left( 1-\sum\limits_{j=1}^{r}\beta _{q+j}\,L^{j}\sum\limits_{m=0}^{\infty
}\left( \sum\limits_{i=1}^{q}\beta _{i}L^{i}\right) ^{m}\right) y_{t}=\frac{1%
}{1-\sum\limits_{i=1}^{q}\beta _{i}}\beta _{0}.
\end{equation*}%
Now we obtain the autoregressive polynomial $g(x)$ of $y_{t}$ which is
\begin{equation}
g(x):=1-\sum\limits_{j=1}^{r}\beta _{q+j}\,x^{j}\sum\limits_{m=0}^{\infty
}\left( \sum\limits_{i=1}^{q}\beta _{i}x^{i}\right) ^{m}.
\end{equation}%
So the second necessary condition for $\left\{ y_{t}\right\} $ to be nonexplosive is
that the roots of $g(x)$ are outside the unit circle.
When there exists at least one $\beta_i \neq 0, i\in\{1,\ldots,q\}$, this second condition is equivalent to require that the roots of $g_{1}(x):=1-\sum\limits_{i=1}^{q}\beta _{i}x^{i}-\sum\limits_{j=1}^{r}\beta_{q+j}\,x^{j}$ and the common roots of $g_{2}(x):=1 - \sum\limits_{i=1}^{q}\beta _{i}x^{i}$ and $g_{3}(x):= 1 - \sum\limits_{j=1}^{r}\beta_{q+j}\,x^{j}$ all are outside the unit circle. More examples of CAViaR DGPs are illustrated in Appendix~\ref{sec:appendix_figs}, among which we can find explosive DGPs which break the condition on the roots of $g_1(x)$ but meet the condition on the common roots of $g_{2}(x)$ and $g_{3}(x)$.

We can review Figure~\ref{fig:ts_simulations_1} with the above stability conditions. CAViaR DGP 1.b, 1.c, 2.b and 2.c meet the above conditions and we also see their nonexplosive behaviours in the plots. The nonexplosiveness of CAViaR DGP 2.a can also be ensured since it has a narrower spread in theory in comparison with CAViaR DGP 2.b. On the other hand, we know that CAViaR DGP 1.a has a downward trend due to the negative term $-0.5|y_{t-1}|$ and hence is explosive.

\subsection{Estimation algorithm}
\label{sec:CAViaR_Algorithm} 
The estimation for CAViaR models can be achieved by the differential evolutionary genetic algorithm~\citep{storn1997differential} used by \cite{engle2004caviar}. Suppose the model $f_{t}(\boldsymbol{\beta })$ is specified as~\eqref{eq:CAViaR} for data $\left\{ y_{t}\right\} _{t=1}^{T}$.
We want to obtain the parameter estimator $\widehat{\boldsymbol{\beta }}$ by the following optimization:
\begin{equation}
\left\{
\begin{aligned}
\widehat{\boldsymbol{\beta }}
			& =\argmin_{\bm{\beta}\in \mathbb{R}^{p+1}} S_T(\bm{\beta})
			\\
			S_T(\bm{\beta})
			& := \sum\limits_{t=1}^{T}\rho _{\tau}\left( y_{t}-f_{t}(\boldsymbol{\beta })\right) 
\end{aligned}
\right.
\label{eq:estimating_beta_argmin}
\end{equation}%
where $S_T(\bm{\beta})$ is the objective function in quantile regressions, and $\rho _{\tau }(x):=x\left( \tau -\bm{1}\{x<0\}\right) $ is called check function~\citep{koenker2005quantile} with the indicator function $\bm{1}\{\cdot \}$.

Following the steps below, we can obtain $\widehat{\boldsymbol{\beta }}$ in~\eqref{eq:estimating_beta_argmin}.

\begin{enumerate}
\item[Step 1:] Generate $n$ (say $10^4$) trial vectors independently from a
uniform distribution $\mathcal{U}(\bm{b}_L,\bm{b}_p)$ as $n$ parameter initial trials, where $\bm{b}_L$ and $\bm{b}_p$ are $(p+1)\times 1$ vectors roughly covering the lower and upper bounds of the true parameter vector $\boldsymbol{\beta
}_{\tau}^o$ of the underlying process in our belief. It is worth mentioning that the values of $\left\{ f_{1-i}(\boldsymbol{\beta}_{\tau}), i = 1,\ldots,q \right\} $ and $\left\{ y_{1-j}, j = 1,\ldots,r \right\} $ acting as initial conditions are
also input-demanded in order to calculate $\{f_{t}(\boldsymbol{\beta })\}_{t=1}^T$ for any $\bm{\beta}\in \mathbb{R}^{p+1}$. For instance, as used
by \cite{engle2004caviar} $f_{0}(\boldsymbol{\beta}_{\tau})$ is given as the estimated $\tau$-th quantle of $\left\{ y_t\right\}_{t=1}^{\lfloor 0.1\,T\rfloor}$ and is fixed in the optimization.\footnote{ $\lfloor\cdot\rfloor$ is known as the floor function (or the greatest integer function) and $\lfloor\cdot\rfloor: \mathbb{R} \to \mathbb{Z}$ of a real number $x$ denotes the greatest integer less than or equal to $x$.} 

\item[Step 2:] Each parameter initial is used to kick off a minimization routine\footnote{ The Nelder Mead simplex algorithm is used in our minimization routine.} on the objective function $S_T(\bm{\beta})$, and the returned value of $\widehat{\boldsymbol{\beta }}$ from the routine and its objective function value are stored.

\item[Step 3:] Select $m$ (say 10) returned vectors of $\widehat{\boldsymbol{\beta }}$ which result in the lowest $m$ values among the $n$ stored objective function values.

\item[Step 4:] Denote the $m$ selected vectors as $\widehat{\boldsymbol{\beta}}^{(1)}, \ldots, \widehat{\boldsymbol{\beta}}^{(m)}$ and use them as initials to restart the minimization routine individually, and update $\widehat{\boldsymbol{%
\beta}}^{(1)}, \ldots, \widehat{\boldsymbol{\beta}}^{(m)}$ with the newly returned vectors respectively.

\item[Step 5:] Repeat Step 4 $a$ (say $5$) times.

\item[Step 6:] Calculate $S_T(\widehat{\boldsymbol{\beta}}^{(i)}), i=1,\ldots,m$. And set the solution to be $\widehat{\boldsymbol{\beta }} = \argmin\limits_{i=1,\ldots,m} S_T(\widehat{\boldsymbol{\beta}}^{(i)})$.
\end{enumerate}

We implement the above estimation algorithm throughout this paper for CAViaR model parameter estimations. There might be a concern if the artificial input of the initial values $\left\{ f_{1-i}(\boldsymbol{\beta}_{\tau}), i = 1,\ldots,q \right\} $ and $\left\{ y_{1-j}, j = 1,\ldots,r \right\} $ affects the parameter estimator. In fact, the effect usually is small and can be neglected when the sample size is large enough because the fitted conditional quantiles $\{f_{t}(\widehat{\bm{\beta }})\}_{t=1}^T$ are kept close to the true ones $\{f_{t}(\bm{\beta })\}_{t=1}^T$ such that it can minimize the objective function despite some burn-in period.

\section{Adaptive random bandwidth method for CAViaR covariance matrix estimation}
\label{sec:covariance_mat_estimation}
Consistency and asymptotic normality of CAViaR model parameters have been proved by~\cite{engle2004caviar}. After regressing data onto a CAViaR model, we would like to implement an inference testing on whether the model is correctly specified. In this section we first investigate how we result in the asymptotic normality of CAViaR model parameter estimators. We focus on the elements of the asymptotic covariance matrix to highlight their roles in connecting sample elements with the corresponding limit behaviours. Next, we check whether existing estimation strategies can perform robustly and satisfactorily for Wald tests on CAViaR models. Finally, we propose a new method called \textit{adaptive random bandwidth} for CAViaR models.

\subsection{Asymptotics of CAViaR}
\label{subsec:derive_asymp_normality}
Consider a time series $\{y_t\}$ of random variables $y_t$ on a complete prbability space $(\Omega, \mathcal{F}, P)$~\footnote{See the assumption C0 of~\cite{engle2004caviar}. We also apply this assumption throughout this paper. That is to say, all the random variables considered in this paper are assumed on a complete prbability space $(\Omega, \mathcal{F}, P)$. }. For applying a generic CAViaR model~\eqref{eq:CAViaR} on $\{y_t\}$, the consistency and asymptotic normality of the estimator $\widehat{\boldsymbol{\beta}}:= \argmin\limits_{\bm{\beta}}\sum\limits_{t=1}^{T} \rho_{\tau}\left( y_t - f_t(\boldsymbol{\beta}) \right)$ has been derived out by~\cite{engle2004caviar}:
\begin{theorem}[Asymptotics given by~\cite{engle2004caviar}] 
\label{thm:AN_CAViaR}
\ \\
For a data generating process $\{ y_t \}$ with its time $t$ conditional $\tau$-th quantile following a generic CAViaR model as~\eqref{eq:CAViaR} parametrized by $\bm{\beta}^{o}$, it satisfies the regularity conditions (C0,\ldots, C7, AN1,\ldots, AN7) in the proof of~\cite{engle2004caviar}. Then
\begin{equation}
\sqrt{T}A_T^{-1/2}D_T \left(\widehat{\bm{\beta}} - \bm{\beta}^{o} \right) \dsim N(\bm{0},\bm{I}_{(p+1)\times(p+1)}),
\label{eq:asym_CAViaR}
\end{equation}
where 
\begin{equation}
\begin{aligned}
\widehat{\boldsymbol{\beta}}
		&
		:= \argmin\limits_{\bm{\beta}\in\mathbb{R}^{p+1}}\sum\limits_{t=1}^{T} \rho_{\tau}\left( y_t - f_t(\boldsymbol{\beta}) \right),
		\\
A_T		& 
		:= \mathbb{E}\left[ T^{-1} \tau(1-\tau)\Sum{T}{t=1} \nabla'f_t(\bm{\beta^{o}} )  \nabla f_t(\bm{\beta^{o}} )\right],
				\\
D_T		& 
		:= \mathbb{E}\left[ T^{-1} \Sum{T}{t=1} h_t\left(0 \middle| \mathcal{F}_{t-1} \right) \nabla'f_t(\bm{\beta^{o}} )  \nabla f_t(\bm{\beta^{o}} )\right],	
				\\
\epsilon_{\tau\,t}
		&
		:= y_t - f_t(\bm{\beta}^{o}),	
\end{aligned}
\label{eq:A&D}
\end{equation}
and $h_t(0| \mathcal{F}_{t-1})$ is denoted as the probability density of $\epsilon_{\tau\,t}$ evaluated at $0$ conditional on the information set $\mathcal{F}_{t-1}$.
$\bm{I}_{(p+1)\times(p+1)}$ is the $(p+1)\times(p+1)$ identity matrix.
\end{theorem}

The above theorem is useful for quantile model (mis)specification tests. For instance, Wald tests can be used to check whether the current model is correctly specified by testing the validity of a more parsimonious nested model. To perform such a quantile model specification test, it often requires to estimate $A_T$, $h_t(0| \mathcal{F}_{t-1})$ and $D_T$. When using traditional estimates $\widehat{A}_T$, $\{\widehat{h}_t(0| \mathcal{F}_{t-1})\}$, $\widehat{D}_T$ of $A_T$, $\{h_t(0| \mathcal{F}_{t-1})\}$ and $D_T$ respectively, we found considerable size distortions in inference tests on CAViaR models in general. We will show that the reason lies in the inaccuracy of $\{\widehat{h}_t(0| \mathcal{F}_{t-1})\}$ in the next subsection. In order to spot the discrepancy in approximating $\{h_t(0| \mathcal{F}_{t-1})\}$, we need a clear picture on how $\left\{ h_{t}(0|\mathcal{F}_{t-1})\right\} $ comes up into the asymptotic normality of the model parameter estimator. Doing so, we can see the role of $\left\{ h_{t}(0|\mathcal{F}_{t-1})\right\} $ and whether a sequence $\left\{ \widehat{h_{t}}(0|\mathcal{F}
_{t-1})\right\} $ is capable to achieve the same role in practice. Let us review the proof of~\cite{engle2004caviar} for Theorem~\ref{thm:AN_CAViaR}
below.

\vspace{0.8cm}

The proof of~\cite{engle2004caviar} is obtained by applying Theorem 3 of~\cite{huber1967behavior} onto $T^{-1/2}\Sum{T}{t=1} \left( \bm{1}\left\{ y_t \leq f_t(\widehat{\boldsymbol{\beta}})  \right\} - \tau\right)\nabla' f_t(\widehat{\boldsymbol{\beta}})$ and the central limit theorem onto $T^{-1/2}\Sum{T}{t=1} \left( \bm{1}\left\{ y_t \leq f_t(\bm{\beta}^{o})  \right\} - \tau\right)\nabla' f_t(\bm{\beta}^{o})$. Huber's conditions are verified in the proof before applying Huber's theorem. 
Denote
\begin{equation}
\begin{aligned}
\text{Hit}_t(\bm{\beta}) 	& := \bm{1}\left\{ y_t \leq f_t(\boldsymbol{\beta})  \right\} - \tau,
							\\
g_t(\boldsymbol{\beta})		& := \nabla f_t(\boldsymbol{\beta}).									
\end{aligned}
\end{equation}
$\text{Hit}_t(\bm{\beta}) $ gives value $-\tau$ every time $y_t$ exceeds $f_t(\boldsymbol{\beta})$ and $1-\tau$ otherwise. With the true underlying parameter $\bm{\beta}^{o}$, $\{ \text{Hit}_t(\bm{\beta}^{o}) \}$ is a martingale difference sequence with respect to $\{\mathcal{F}_{t-1}\}$.   
It is easy to get that $T^{-1/2}\Sum{T}{t=1} \text{Hit}_t(\bm{\beta}^{o}) g_t(\bm{\beta}^{o})$ follows the central limit theorem because $\left\{  \text{Hit}_t(\bm{\beta}^{o}) g_t(\bm{\beta}^{o}) \right\}$ is a martingale difference sequence with the assumption AN1 of~\cite{engle2004caviar} on its uniformly bounded second moment. So we get that
\begin{equation}
T^{-1/2}\Sum{T}{t=1} \text{Hit}_t(\bm{\beta}^{o}) g_t(\bm{\beta}^{o}) \dsim N(0,A_T).
\label{eq:CLT_beta0}
\end{equation}

It has also been proved by~\cite{engle2004caviar} that 
\begin{equation}
T^{-1/2}\Sum{T}{t=1}    \text{Hit}_t(\widehat{\bm{\beta}})g_t(\widehat{\bm{\beta}}) = o_p(1).
\label{eq:op_gradient_of_CF}
\end{equation} 

Next, we are going to manifest $\left\{h_t(0|\mathcal{F}_{t-1})\right\}$ in the proof in a way which makes the appearance of $\left\{h_t(0|\mathcal{F}_{t-1})\right\}$ more intuitive. We rewrite $\text{Hit}_t(\widehat{\bm{\beta}})\, g_t(\widehat{\bm{\beta}})$ as follows:
\begin{equation}
\begin{aligned}
\text{Hit}_t(\widehat{\bm{\beta}})\, g_t(\widehat{\bm{\beta}})\,  		
		& = 
		\left( \text{Hit}_t(\bm{\beta}^{o}) + \text{Hit}_t(\widehat{\bm{\beta}}) - \text{Hit}_t(\bm{\beta}^{o})  \right) g_t(\widehat{\bm{\beta}})
		\\
		& = 
		  \text{Hit}_t(\bm{\beta}^{o})g_t(\widehat{\bm{\beta}})
		 + 
		  \left( \text{Hit}_t(\widehat{\bm{\beta}}) - \text{Hit}_t(\bm{\beta}^{o}) \right) g_t(\widehat{\bm{\beta}}).  
\end{aligned} 
\label{eq:taylor_beta0_on_betahat}
\end{equation}
Take expectation on the both sides of Equation~\eqref{eq:taylor_beta0_on_betahat} and get 
\begin{equation}
\begin{aligned}
& T^{-1/2}\Sum{T}{t=1} \mathbb{E} \left[
	  \text{Hit}_t(\widehat{\bm{\beta}})\, g_t(\widehat{\bm{\beta}})\,  
		\right]
		\\		
		& = 
		T^{-1/2}\Sum{T}{t=1}  \mathbb{E} \left[   \text{Hit}_t(\bm{\beta}^{o})g_t(\widehat{\bm{\beta}})
		 + 
		 \left( \text{Hit}_t(\widehat{\bm{\beta}}) - \text{Hit}_t(\bm{\beta}^{o}) \right) g_t(\widehat{\bm{\beta}})
		 \right]
		\\
		& = 
		 T^{-1/2}\Sum{T}{t=1}  \mathbb{E} \left[\mathbb{E} \left[\text{Hit}_t(\bm{\beta}^{o}) \middle| \mathcal{F}_{t-1} \right] g_t(\widehat{\bm{\beta}}) \right]
		 + 
		T^{-1/2}\Sum{T}{t=1}  \mathbb{E} \left[ \left( \text{Hit}_t(\widehat{\bm{\beta}}) - \text{Hit}_t(\bm{\beta}^{o}) \right) g_t(\widehat{\bm{\beta}})	\right]
		 \\
		 & = 
		 T^{-1/2}\Sum{T}{t=1} \mathbb{E} \left[ 
		 \mathbb{E} \left[\left( \text{Hit}_t(\widehat{\bm{\beta}}) - \text{Hit}_t(\bm{\beta}^{o}) \right) \middle| \mathcal{F}_{t-1} \right]
		 g_t(\bm{\beta}^{o})	\right] 
		 \\
		 & +
		 T^{-1/2}\Sum{T}{t=1}  \mathbb{E} \left[ 
		 \mathbb{E} \left[\left( \text{Hit}_t(\widehat{\bm{\beta}}) - \text{Hit}_t(\bm{\beta}^{o}) \right) \middle| \mathcal{F}_{t-1} \right]
		 \left( g_t(\widehat{\bm{\beta}}) - g_t(\bm{\beta}^{o})	\right)	\right]
		  \\
		 & = 
		 T^{-1/2}\Sum{T}{t=1} \mathbb{E} \left[ 
		 \mathbb{E} \left[\left( \text{Hit}_t(\widehat{\bm{\beta}}) - \text{Hit}_t(\bm{\beta}^{o}) \right) \middle| \mathcal{F}_{t-1} \right]
		 g_t(\bm{\beta}^{o})	\right]		 
		 \\
		 & +
		 T^{-1/2}\Sum{T}{t=1}  \mathbb{E} \left[ 
		 \mathbb{E} \left[\left( \text{Hit}_t(\widehat{\bm{\beta}}) - \text{Hit}_t(\bm{\beta}^{o}) \right) \middle| \mathcal{F}_{t-1} \right] \right] \mathcal{O}_p( \norm{ \widehat{\bm{\beta}} - \bm{\beta}^{o} }_{\infty} ),
\end{aligned} 
\label{eq:expectation_grad_OB_QR}
\end{equation}
where $\norm{\cdot}_{\infty}$ is the supremum norm of vectors.
And
\begin{equation}
\begin{aligned}
	& \mathbb{E} \left[\left( \text{Hit}_t(\widehat{\bm{\beta}}) - \text{Hit}_t(\bm{\beta}^{o}) \right) \middle| \mathcal{F}_{t-1} \right] 
	 = 
	\text{P}\left\{ y_t \leq f_t(\widehat{\bm{\beta}}) \middle| \mathcal{F}_{t-1}  \right\}
	-
	\text{P}\left\{ y_t \leq f_t(\bm{\beta}^{o}) \middle| \mathcal{F}_{t-1}  \right\} 
	\\
	& =
	F_t\left(f_t(\widehat{\bm{\beta}}) \middle| \mathcal{F}_{t-1}  \right)
	-
	F_t\left( f_t(\bm{\beta}^{o}) \middle| \mathcal{F}_{t-1} \right)
	\\
	& = 
	F_t'\left( f_t(\bm{\beta}^{o} ) \middle| \mathcal{F}_{t-1} \right)\left(  f_t(\widehat{\bm{\beta}}) - f_t(\bm{\beta}^{o}) \right) 
	+ 
	\mathcal{O}_p\left(  f_t(\widehat{\bm{\beta}}) - f_t(\bm{\beta}^{o}) \right)^2    
	\\
	& = 
	h_t(0|\mathcal{F}_{t-1})\left(  \left( \widehat{\bm{\beta}} - \bm{\beta}^{o} \right)' \nabla f_t(\bm{\beta}^{o})
	+ 
	\mathcal{O}_p( \norm{ \widehat{\bm{\beta}} - \bm{\beta}^{o} }^2_{\infty} )  \right) 
	 +
	\mathcal{O}_p( \norm{ \widehat{\bm{\beta}} - \bm{\beta}^{o} }^2_{\infty} )   
	\\
	& = 
	 \left( \widehat{\bm{\beta}} - \bm{\beta}^{o} \right)' h_t(0|\mathcal{F}_{t-1}) \nabla f_t(\bm{\beta}^{o}) 
	 + 
	 \mathcal{O}_p( \norm{ \widehat{\bm{\beta}} - \bm{\beta}^{o} }^2_{\infty} ),
\end{aligned}	
\label{eq:expectation_on_hithat_hit0}
\end{equation}
where $F_t\left( \cdot\middle| \mathcal{F}_{t-1} \right) $ is the probability density function of $y_t$ conditional on $\mathcal{F}_{t-1}$, and $ h_t(0|\mathcal{F}_{t-1})= F_t'\left( f_t(\bm{\beta}^{o} ) \middle| \mathcal{F}_{t-1} \right) $.
Substituting~\eqref{eq:expectation_on_hithat_hit0} into~\eqref{eq:expectation_grad_OB_QR} gives
\begin{equation}
\begin{aligned}
& T^{-1/2}\Sum{T}{t=1} \mathbb{E} \left[
	  \text{Hit}_t(\widehat{\bm{\beta}})\, g_t(\widehat{\bm{\beta}})\,  
		\right]
		 \\
		 & = 
		  \left( \widehat{\bm{\beta}} - \bm{\beta}^{o} \right)'\cdot T^{-1/2}\Sum{T}{t=1} \mathbb{E} \left[ 
		 h_t(0|\mathcal{F}_{t-1}) \nabla f_t(\bm{\beta}^{o}) \nabla' f_t(\bm{\beta}^{o}) 
		\right] 
		 \\
		 & +
		  \mathcal{O}_p( T^{1/2}\norm{ \widehat{\bm{\beta}} - \bm{\beta}^{o} }^2_{\infty} ) .
\end{aligned} 
\label{eq:expectation_grad_OB_QR_final}
\end{equation}
\\
Success in applying Huber's theorem gives
\begin{equation}
		T^{-1/2}\Sum{T}{t=1} \mathbb{E} \left[
	  	\text{Hit}_t(\widehat{\bm{\beta}})\, g_t(\widehat{\bm{\beta}})\,  
		\right] 
		=
		-T^{-1/2}\Sum{T}{t=1} \left( \bm{1}\left\{ y_t \leq f_t(\bm{\beta}^{o})  \right\} - \tau\right)\nabla' f_t(\bm{\beta}^{o})
		+ 
		\mathit{o}_p(1)
\label{eq:huber_theorem_equality}
\end{equation}
Therefore, the asymptotic normality of $ T^{1/2}\left(\widehat{\bm{\beta}} - \bm{\beta}^{o} \right)$ is obtained by substituting~\eqref{eq:CLT_beta0} and \eqref{eq:expectation_grad_OB_QR_final} into~\eqref{eq:huber_theorem_equality}.

\vspace{0.8cm}
\noindent From the above derivation, it is clear that the role of $ h_t(0|\mathcal{F}_{t-1})$ is actually an approximation to $F_t'\left( f_t(\bar{\bm{\beta}} ) \middle| \mathcal{F}_{t-1} \right)$ in which $\bar{\bm{\beta}}$ is between $\bm{\beta}^{o}$ and $\widehat{\bm{\beta}}$. This role comes to the surface of~\eqref{eq:expectation_on_hithat_hit0} using the fact that
\begin{equation}
F_t\left( f_t(\widehat{\bm{\beta}})  \middle| \mathcal{F}_{t-1} \right) - F_t\left( f_t(\bm{\beta}^{o})  \middle| \mathcal{F}_{t-1} \right) = F_t'\left( f_t(\bar{\bm{\beta}} ) \middle| \mathcal{F}_{t-1} \right) \left( \nabla'f_t(\bm{\beta}^{o}) \left( \widehat{\bm{\beta}} - \bm{\beta}^{o} \right) \right)
\label{eq:ht_role}
\end{equation} 
by the Mean Value Theorem. This approximating role of $h_t(0|\mathcal{F}_{t-1})$ sets a clear mission of any $\widehat{h_t}(0|\mathcal{F}_{t-1})$ supposed to achieve, which can be used to examine an estimator for $h_t(0|\mathcal{F}_{t-1})$ as well as to propose an improved estimation method. In next subsection, we are going to examine the performances of some existing methods for estimating $h_t(0|\mathcal{F}_{t-1})$ and the role of $h_t(0|\mathcal{F}_{t-1})$ will help to find out the intrinsic defects of those methods.

\subsection{Existing methods for CAViaR covariance matrix estimation }
Based on the literature on quantile regressions, in general there are two ways to estimate $\left\{ h_t\left(0 \middle| \mathcal{F}_{t-1} \right) \right\}$ in $D_T$ with $\{ \epsilon_{\tau\,t}\}$ being potentially non-i.i.d.. One is referred to as the \textit{Hendricks Koenker Sandwich Approach}~\citep{hendricks1992hierarchical, koenker2005quantile} analogous to the \textit{finite difference} idea resulting in the estimator $\widehat{h_t}^{fd}(0 | \mathcal{F}_{t-1} ) $ for $h_t(0 | \mathcal{F}_{t-1} )$ as follows: 
\begin{equation}
\widehat{h_t}^{fd}\left(0 \middle| \mathcal{F}_{t-1} \right) = \frac{2\,\Delta \tau_T}{ f_{t}(\boldsymbol{\beta}_{\tau + \Delta\tau_T}) -  f_{t}(\boldsymbol{\beta}_{\tau - \Delta\tau_T}) },
\label{eq:finite_diffrence_snadwich}
\end{equation}  
where $\Delta\tau_T$ is subject to $ 0<\tau \pm \Delta\tau_T <1$ with $\Delta\tau_T \rightarrow 0$ as $T \rightarrow \infty$. The other one is referred to as the \textit{Powell Sandwich}~\citep{powell1991estimation, koenker2005quantile} based on the kernel density estimation idea resulting in the estimator $\widehat{h_t}^{kernel}(0 | \mathcal{F}_{t-1} ) $ for $h_t(0 | \mathcal{F}_{t-1} )$ as follows: 
\begin{equation}
\begin{aligned}
\widehat{h_t}^{kernel}\left(0 \middle| \mathcal{F}_{t-1} \right) 
			& 	= 
			\frac{ \text{P}\{ y_t \leq f_{t}(\boldsymbol{\beta}_{\tau}) + c_T | \mathcal{F}_{t-1} \}   - \text{P}\{ y_t \leq f_{t}(\boldsymbol{\beta}_{\tau}) - c_T| \mathcal{F}_{t-1}\}}{ 2\,c_T }
			\\
			& \approx \frac{1}{2\,c_T}\,K\left( \frac{ y_t - f_{t}(\boldsymbol{\beta}_{\tau})}{ 2\,c_T } \right)
\end{aligned}
\label{eq:powell_sandwich}
\end{equation} 
where $K(\cdot)$ is a suitable kernel function with bandwidth $2\,c_T$ and $c_T \rightarrow 0$ as $T \rightarrow \infty$. As we can see in~\eqref{eq:powell_sandwich}, one kernel function is applied throughout $\{y_t\}$ with $ y_t - f_{t}(\boldsymbol{\beta}_{\tau})$ being the only distinguishable information for $\widehat{h_t}^{kernel}(0 | \mathcal{F}_{t-1} ) $. Therefore, this kernel method does not capture sufficient information to distinguish  time-varying conditional distributions of $\{y_t\}$, and consequently cannot fully adapt to the time-variations. Additionally, the choice of the kernel function $K(\cdot)$ and the bandwidth parameter $c_T$ are still in a lot of nettlesome questions in practice. A similar issue in the \textit{Hendricks Koenker Sandwich Approach} is on choosing $\Delta\tau_T $ and extra error resulted from estimating $ f_{t}(\boldsymbol{\beta}_{\tau + \Delta\tau_T}) $ and $ f_{t}(\boldsymbol{\beta}_{\tau - \Delta\tau_T}) $. 

The estimation method adopted by~\cite{engle2004caviar} is a form of the \textit{Powell Sandwich} as follows:
\begin{equation}
\widehat{h_t}^{ker}\left(0 \middle| \mathcal{F}_{t-1} \right) = \frac{ \bm{1}\{ | y_t - f_t(\widehat{\bm{\beta}}_{\tau} ) | < \widehat{c}_T \}  }{ 2\,c_T }
\label{eq:parzon_window}
\end{equation}
As suggested by~\cite{koenker2005quantile} and~\cite{machado2013quantile}, the bandwidth $\widehat{c}_T$ generally adopted is defined as follows:
\begin{equation}
\widehat{c}_T = \widehat{k}_T \left[ \Phi^{-1}(\tau + m_T) - \Phi^{-1}(\tau - m_T) \right],
\label{eq:cT}
\end{equation} 
where $m_T$ is defined as 
\begin{equation}
\widehat{m}_T = T^{-\frac{1}{3}}\left( \Phi^{-1}(1 - \frac{0.05}{2})  \right)^{\frac{2}{3}}\, \left( \frac{1.5 (\phi(\Phi^{-1}(\tau)))^2}{2(\Phi^{-1}(\tau))^2 + 1}  \right)^{\frac{1}{3}},
\end{equation} 
with $\Phi(\cdot)$ and $\phi(\cdot)$ being the cumulative distribution and probability density functions of $N(0,1)$ respectively. And $\widehat{k}_T $ is defined as the median absolute deviation of the conditional $\tau$-th quantile regression residuals.

Wald tests are applied in this subsection to check the performances of the above estimation methods for CAViaR models.

First, we consider the following candidate model specifications for the conditional $\tau$-th ($\tau\in (0,1)$) quantile of a time series $\left\{ y_t \right\}$ with $f_t(\boldsymbol{\beta}_{\tau})$ denoted as the $\tau$-th quantile of $y_t$ conditional on the information set $\mathcal{F}_{t-1}$.
\[
\left\{
\begin{tabular}{p{.8\textwidth}}
\begin{itemize}
\item full specification:
\begin{equation}
f_t(\boldsymbol{\beta}_{\tau}^{FM}) = \beta_0^{FM}(\tau) + \beta_1^{FM} f_{t-1}(\boldsymbol{\beta}_{\tau}^{FM}) + \beta_{2}^{FM}\, \left( y_{t-1} \right)^{+} + \beta_{3}^{FM}\, \left( y_{t-1} \right)^{-},
\label{eq:CAViaR_Full_AS_Wald}
\end{equation}
where the operators $(\cdot)^{+}$ and $(\cdot)^{-}$ are defined as $(x)^{+} = \max(x,0), (x)^{-} = -\min(x,0)$.

\item restrictive model 1:
\begin{equation}
\begin{aligned}
f_t(\bm{\beta}_{\tau}^{R1}) & = \beta_0(\tau)^{R1} + \beta_1^{R1} f_{t-1}(\boldsymbol{\beta}_{\tau}^{R1}) + \beta_{2}^{R1}\, | y_{t-1} |
								\\
								& = \beta_0(\tau)^{R1} + \beta_1^{R1} f_{t-1}(\boldsymbol{\beta}_{\tau}^{R1}) + \beta_{2}^{R1}\, \left( y_{t-1} \right)^{+} + \beta_{2}^{R1}\, \left( y_{t-1} \right)^{-}.
\end{aligned}
\label{eq:CAViaR_RES_abs}
\end{equation}

\item restrictive model 2:
\begin{equation}
\begin{aligned}
f_t(\boldsymbol{\beta}_{\tau}^{R2}) 	& = \beta_0^{R2}(\tau) + \beta_1^{R2} f_{t-1}(\boldsymbol{\beta}_{\tau}^{R2}) + \beta_{2}^{R2}\, y_{t-1}
								\\
								& = \beta_0^{R2}(\tau) + \beta_1^{R2} f_{t-1}(\boldsymbol{\beta}_{\tau}^{R2}) + \beta_{2}^{R2}\,\left( y_{t-1} \right)^{+} - \beta_{2}^{R2}\, \left( y_{t-1} \right)^{-}.
\end{aligned}
\label{eq:CAViaR_RES_Y}
\end{equation}

%

\end{itemize}
\end{tabular}
\right.
\]

The models~\eqref{eq:CAViaR_RES_abs} and \eqref{eq:CAViaR_RES_Y} are nested within model~\eqref{eq:CAViaR_Full_AS_Wald}. 
Now let us consider the Wald test on models~\eqref{eq:CAViaR_Full_AS_Wald} and \eqref{eq:CAViaR_RES_abs} first. Simulate a time series $\left\{ y_t \right\}$ with its DGP specified as the model~\eqref{eq:CAViaR_RES_abs} with the underlying parameter vector $\bm{\beta}_{u_t}^{R1} = [F^{-1}_{N(0,1)}(u_t), 0.2, 0.3]'$, where $\left\{ u_t \right\}\overset{i.i.d.}{\sim} \mathcal{U}(0,1)$ and $F^{-1}_{N(0,1)}(\cdot)$ is the inverse standard normal probability distribution function. The sample size of each simulated sample is $4000$. Conditional $50\%$-th quantiles are estimated for each of total $1000$ simulated samples in this DGP by regressing the sample onto the full model~\eqref{eq:CAViaR_Full_AS_Wald}. The Wald test implemented here consists of the null hypothesis of the form $\mathit{H_0}: R\bm{\beta}_{\tau}^{FM}= \gamma$, where $R = [ 0 , 0, 1, -1] $, $\gamma = 0$, and $\widehat{\bm{\beta}}_{\tau}$ is the estimator of the full model parameter vector in~\eqref{eq:CAViaR_Full_AS_Wald}. The Wald test statistic denoted by $W_T$ is formulated~\citep{weiss1991estimating} as follows:
\begin{equation}
\label{eq:wald_test_stat}
W_T = T \left( R\widehat{\bm{\beta}}_{\tau} - \gamma\right)'\left[ R \widehat{D}^{-1}_T \widehat{A}_T \widehat{D}^{-1}_T R'\right]^{-1}\left( R\widehat{\bm{\beta}}_{\tau} - \gamma\right),
\end{equation}
where $\widehat{A}_T$ and $\widehat{D}_T$ are estimates for $A_T$ and $D_T$ in~\eqref{eq:A&D} respectively. It is straightforward to obtain $\widehat{A}_T$ and $\widehat{D}_T$ by plugging in $\widehat{\bm{\beta}}_{\tau}$ and $\left\{ \widehat{h_t}\left(0 \middle| \mathcal{F}_{t-1} \right) \right\}$, i.e.,
$$
\left\{
\begin{aligned}
\widehat{A}_T		
		&	
		=  T^{-1} \tau(1-\tau)\Sum{T}{t=1} \nabla'f_t( \widehat{\bm{\beta}}_{\tau} )  \nabla f_t( \widehat{\bm{\beta}}_{\tau} ),
		\\
\widehat{D}_T		
		&
		=  T^{-1} \Sum{T}{t=1} \widehat{h_t}\left(0 \middle| \mathcal{F}_{t-1} \right) \nabla'f_t(\widehat{\bm{\beta}}_{\tau} )  \nabla f_t(\widehat{\bm{\beta}}_{\tau} ).
\end{aligned}
\right.
$$

Notations on $\widehat{D}_T$ to distinguish different estimators used for $\left\{ h_t\left(0 \middle| \mathcal{F}_{t-1} \right) \right\}$ are given by
\begin{equation}
\widehat{D}_T^{ker}	 = \left( 2 T \widehat{c}_T \right)^{-1} \Sum{T}{t=1}\bm{1}\{ | y_t - f_t(\widehat{\bm{\beta}}_{\tau} ) | < \widehat{c}_T \} \, \nabla'f_t(\widehat{\bm{\beta}}_{\tau} )  \nabla f_t(\widehat{\bm{\beta}}_{\tau} ),	
\label{eq:hat_D_fixed_bandwidth}
\end{equation}
\begin{equation}
\widehat{D}_T^{fd}	 = \left(  T  \right)^{-1} \Sum{T}{t=1} \frac{2\,\Delta \tau_T}{ f_{t}(\boldsymbol{\beta}_{\tau + \Delta\tau_T}) -  f_{t}(\boldsymbol{\beta}_{\tau - \Delta\tau_T}) } \, \nabla'f_t(\widehat{\bm{\beta}}_{\tau} )  \nabla f_t(\widehat{\bm{\beta}}_{\tau} ),	
\label{eq:hat_D_finite_difference}
\end{equation}
where $\widehat{c}_T$ is determined as~\eqref{eq:cT}.

We are going to examine each element in the estimation of $D_T$. The analytic  solution to $ h_t\left(0 \middle| \mathcal{F}_{t-1} \right)$ can be obtained as follows:
\begin{equation}
\begin{aligned}
h_t\left(0 \middle| \mathcal{F}_{t-1} \right)
					& = \frac{\partial \tau}{\partial f_t(\bm{\beta_{\tau}})}
					 = \frac{1}{\frac{\partial \beta_0(\tau)}{\partial \tau} + \beta_1 \frac{\partial f_{t-1}(\bm{\beta_{\tau}})}{\partial \tau}}
					\\
					& = \frac{1}{ 
					\frac{\partial \beta_0(\tau)}{\partial \tau}\Sum{n}{i=0} \beta_1^i + \beta_1^{n+1} \frac{\partial f_{t-n-1}(\bm{\beta_{\tau}})}{\partial \tau}
					}
					\\
					& = \left( 1 - \beta_1 \right) \frac{1}{ \beta_0'(\tau)}
\end{aligned}
\end{equation}
where $\beta_0'(\tau):=\frac{\partial \beta_0(\tau)}{\partial \tau}$. The last line is obtained by knowing $\abs{\beta_1} < 1$.  The analytic  solution to $ h_t\left(0 \middle| \mathcal{F}_{t-1} \right)$ is used to help identify inaccurate elements in $\widehat{D}_T$ by comparing the test performances of using $\widehat{D}_T^{ker}$, $\widehat{D}_T^{fd}$ and the following
\begin{equation}
\widehat{D}_T^{h_0}	 = \left(  T  \right)^{-1} \Sum{T}{t=1} \left( 1 - \beta_1 \right) \frac{1}{ \beta_0'(\tau)} \, \nabla'f_t(\widehat{\bm{\beta}}_{\tau} )  \nabla f_t(\widehat{\bm{\beta}}_{\tau} ).	
\label{eq:hat_D_trueht}
\end{equation} 
The test performances of using $\widehat{D}_T^{ker}$, $\widehat{D}_T^{fd}$ and $\widehat{D}_T^{h_0}$ are shown in Table~\ref{tab:WT_ResABs} and~\ref{tab:WT_ResY}, which are compared together with the Wald test result using the true underlying parameter vector $\bm{\beta}^{FM}_{\tau} = [F^{-1}_{N(0,1)}(\tau), 0.2, 0.3]', \tau\in(0,1)$ into
\begin{equation}
\begin{aligned}
\widehat{D}_T^{0}	
			&  = \left(  T  \right)^{-1} \Sum{T}{t=1} \left( 1 - \beta_1 \right) \frac{1}{ \beta_0'(\tau)} \, \nabla'f_t(\bm{\beta}_{\tau} )  \nabla f_t(\bm{\beta}_{\tau} )
			\\
			&  = \frac{ \left( 1 - 0.2 \right)\, \phi(F^{-1}_{N(0,1)}(\tau))}{T} \Sum{T}{t=1}  \, \nabla'f_t(\bm{\beta}_{\tau}^{o} )  \nabla f_t(\bm{\beta}_{\tau}^{o} ),
\end{aligned}	
\label{eq:hat_D_alltrue}
\end{equation} 
where $\phi(\cdot)$ is the probability density function of $N(0,1)$.

The size performances of the Wald tests on the models~\eqref{eq:CAViaR_Full_AS_Wald} and \eqref{eq:CAViaR_RES_abs} using different $D_T$ estimators are listed in Table~\ref{tab:WT_ResABs} in which each estimated size is obtained by the percentage rejection rate among the 1000 samples of $T=4000$ in the DGP~\eqref{eq:CAViaR_RES_abs}. Analogously, we implement the Wald test on models ~\eqref{eq:CAViaR_Full_AS_Wald} and \eqref{eq:CAViaR_RES_Y} with the underlying DGP $\left\{ y_t \right\}$ specified as the model~\eqref{eq:CAViaR_RES_Y} with the underlying parameter vector $\bm{\beta}_{u_t}^{R2} = [F^{-1}_{N(0,1)}(u_t), 0.2, 0.3]'$, where $\left\{ u_t \right\}\overset{i.i.d.}{\sim} \mathcal{U}(0,1)$. The number of observations in each stimulated sample from this DGP is $4000$. Conditional $50\%$-th quantiles are estimated for each of $1000$ simulated samples by regressing the sample onto the full model~\eqref{eq:CAViaR_Full_AS_Wald}. The Wald test implemented in this case consists of the null hypothesis of the form $\mathit{H_0}: R\bm{\beta}_{\tau}^{FM}= \gamma$, where $R = [ 0 , 0, 1, 1] $, $\gamma = 0$, and $\widehat{\bm{\beta}}_{\tau}$ is the estimator of the full model regression~\eqref{eq:CAViaR_Full_AS_Wald}. In result, the size performances of the Wald tests on \eqref{eq:CAViaR_Full_AS_Wald} and \eqref{eq:CAViaR_RES_Y} are listed in Table~\ref{tab:WT_ResY}.

From Table~\ref{tab:WT_ResABs} and~\ref{tab:WT_ResY}, we can see large size distortions with $\widehat{D}_T^{fd}$, unlike $\widehat{D}_T^{ker}$, $\widehat{D}_T^{h_0}$ or $\widehat{D}_T^{0}$ that are performing in line with the nominal size. This comparison points out the crucial element estimation to the accuracy of $\widehat{D}_T$ which is $\left\{ \widehat{h_t}\left(0 \middle| \mathcal{F}_{t-1} \right) \right\}$. To check whether $\{ \widehat{h_t}^{ker}\left(0 \middle| \mathcal{F}_{t-1} \right)\}$ is capable to achieve the role of $\{ h_t\left(0 \middle| \mathcal{F}_{t-1} \right) \}$ robustly for time-varying conditional probability densities, we consider the following DGP:

\begin{align}
y_t = f_t(\boldsymbol{\beta}_{u_t}^{R3}) 
												& = \beta_0^{R3}(u_t)\sqrt{ \left( y_{t-1} \right)^{+}} + \beta_1^{R3} f_{t-1}(\boldsymbol{\beta}_{\tau}^{R3}) + \beta_{2}^{R3}\, | y_{t-1} |
		\label{eq:timevarying_ht_CAViaR_DGP_b0Bposy_restricted}
				\\
				& = \beta_0^{R3}(u_t)\sqrt{ \left( y_{t-1} \right)^{+}} + \beta_1^{R3} f_{t-1}(\boldsymbol{\beta}_{\tau}^{R3}) + \beta_{2}^{R3}\, \left( y_{t-1} \right)^{+} + \beta_{2}^{R3}\, \left( y_{t-1} \right)^{-},
				\label{eq:timevarying_ht_CAViaR_DGP_b0Bposy}
\end{align}
where $\left\{ u_t \right\}\overset{i.i.d.}{\sim} \mathcal{U}(0,1)$ and the underlying parameters are given as $\bm{\beta}_{u_t}^{R3} = [F^{-1}_{N(0,1)}(u_t), 0.2, 0.3]'$. The analytic form of the corresponding conditional probability density $ h_t\left(0 \middle| \mathcal{F}_{t-1} \right)$ of $y_t$ at its $\tau$-th quantile $ f_t(\bm{\beta}_{\tau})$ given $\mathcal{F}_{t-1}$ can be derived out as follows:
\begin{equation}
\begin{aligned}
h_t\left(0 \middle| \mathcal{F}_{t-1} \right) 
		= \left(\frac{\partial f_t(\boldsymbol{\beta}_{\tau})}{\partial \tau} \right)^{-1} 
		& = \left( \frac{\partial \beta_0(\tau)}{\partial \tau}\sqrt{ \left( y_{t-1} \right)^{+}}  + \beta_1 \frac{\partial f_{t-1}(\boldsymbol{\beta}_{u_t})}{\partial \tau} \right)^{-1} 
		\\
		& = \left(\frac{\partial \beta_0(\tau)}{\partial \tau}\Sum{\infty}{i=1}\beta_1^{i-1}\sqrt{ \left( y_{t-i} \right)^{+}}  \right)^{-1},
\end{aligned}
\label{eq:time_varying_ht_DGP_b0Nposy}
\end{equation}
where the first equation is obtained by iteratively rewriting $ \frac{\partial f_{t-i}(\boldsymbol{\beta}_{\tau})}{\partial \tau} $ at each $i$ and knowing $|\beta_1|<1$. This analytic form of $h_t\left(0 \middle| \mathcal{F}_{t-1} \right) $ in~\eqref{eq:timevarying_ht_CAViaR_DGP_b0Bposy_restricted} shows that $\{h_t\left(0 \middle| \mathcal{F}_{t-1} \right) \}$ indeed is time-varying and nonzero with probability one.

We simulate $1000$ samples from the DGP~\eqref{eq:timevarying_ht_CAViaR_DGP_b0Bposy_restricted} with $T=5000$, and estimate the conditional $50\%$-th quantiles of each sample by regressing the sample onto the full model specification~\eqref{eq:timevarying_ht_CAViaR_DGP_b0Bposy}. The Wald test described as~\eqref{eq:wald_test_stat} with $R = [ 0 , 0, 1, -1] $ is performed on these $1000$ samples and the size performance is presented in Table~\ref{tab:WT_ResABs_DGP_b0Nposy}. We see a large size distortion with the kernel method $\widehat{D}_T^{ker}$ in Table~\ref{tab:WT_ResABs_DGP_b0Nposy}. More tests are conducted for different DGPs and together with the results are presented in Appendix~\ref{sec:appendix_tests}. Based on our test results, we see that the kernel method for estimating $\left\{ h_t\left(0 \middle| \mathcal{F}_{t-1} \right) \right\}$ is not robust and cannot fully adapt to time-varying conditional probability densities.

Estimating $\left\{ h_t\left(0 \middle| \mathcal{F}_{t-1} \right) \right\}$ robustly has to be achieved in order to ensure the reliability of CAViaR analysis based on the asymptotic properties of CAViaR model parameter estimators. In seeking for improving the accuracy of $\left\{ \widehat{h_t}\left(0 \middle| \mathcal{F}_{t-1} \right) \right\}$, we bear in mind two guidances. One is the role of $\left\{ h_t\left(0 \middle| \mathcal{F}_{t-1} \right) \right\}$ on how it links sample elements with the corresponding limit behaviours, see Section~\ref{subsec:derive_asymp_normality}. The other guidance is the fundamental flaws of $\{ \widehat{h_t}^{ker}\left(0 \middle| \mathcal{F}_{t-1} \right)\}$ and $\{ \widehat{h_t}^{fd}\left(0 \middle| \mathcal{F}_{t-1} \right)\}$ in their accuracy. In terms of $\{ \widehat{h_t}^{fd}\left(0 \middle| \mathcal{F}_{t-1} \right)\}$, $\Delta\tau_T$ needs to be determined properly and two more quantile regressions need to be preformed in order to obtain $\widehat{\bm{\beta}}_{\tau + \Delta \tau_T} $ and $\widehat{\bm{\beta}}_{\tau - \Delta \tau_T}$. The effect of this extra estimation error is crucial to the performance of $\{ \widehat{h_t}^{fd}\left(0 \middle| \mathcal{F}_{t-1} \right)\}$. Although $\{ \widehat{h_t}^{ker}\left(0 \middle| \mathcal{F}_{t-1} \right)\}$ does not need extra quantile regressions, it still requires a proper choice on the kernel function~$K(\cdot)$ and the bandwidth $\widehat{c}_T$. Remarkably, $\{ \widehat{h_t}^{ker}\left(0 \middle| \mathcal{F}_{t-1} \right)\}$ does not differentiate the observations within the bandwidth regardless of the number of the observations in the bandwidth while using the kernel function $\bm{1}\{ | y_t - f_t(\widehat{\bm{\beta}}_{\tau} ) | < \widehat{c}_T \} $. Therefore, it is desirable to get rid of choosing bandwidth $\Delta\tau_T$ or $c_T$ and the kernel function~$K(\cdot)$ in the estimation. In the next subsection, a robust estimation method for $\left\{ h_t\left(0 \middle| \mathcal{F}_{t-1} \right) \right\}$ is developed up without the need in choosing a bandwidth or a kernel function.

\subsection{Adaptive random bandwidth method}
We have noticed that the accuracy of the $\left\{ h_t(0|\mathcal{F}_{t-1}) \right\}$ estimation is crucial to the performance of inference tests based on the asymptotic normality of CAViaR model parameter estimators. It is also well known that $\{ \widehat{h_t}^{fd}\left(0 \middle| \mathcal{F}_{t-1} \right)\}$ suffers both from the error in estimating $f_t(\bm{\beta}_{\tau + \Delta\tau})$ and $f_t(\bm{\beta}_{\tau - \Delta\tau})$ and from choosing a proper $\Delta\tau_T$. On the other hand, $\{ \widehat{h_t}^{ker}\left(0 \middle| \mathcal{F}_{t-1} \right)\}$ has some fundamental problems. First of all, $\{ \widehat{h_t}^{ker}\left(0 \middle| \mathcal{F}_{t-1} \right)\}$ cannot fully adapt to time-varying conditional distributions of time series due to the fact that the same kernel function $K(\cdot)$ and only timely information $(y_t - f_t)$ are used in estimating $h_t(0|\mathcal{F}_{t-1})$ for all t. Second, finding a proper kernel function $K(\cdot)$ with a proper bandwidth $c_T$ still faces a lot nettlesome problems in practice. Neither of these two methods is practically robust. The goal in this subsection is to develop an estimation method for  $\left\{ h_t(0|\mathcal{F}_{t-1}) \right\}$ which can adapt to time-variation characteristics of CAViaR DGPs and is robust in practice without the need to determine a proper bandwidth. We name this estimation method as the \textit{adaptive random bandwidth} (ARB) method which can reliably bridge asymptotic properties of CAViaR models in theory with CAViaR applications.

The idea of this method is inspired by viewing the role of $\left\{ h_t \left(0 \middle| \mathcal{F}_{t-1} \right) \right\}$ on how it links sample elements with the corresponding limit behaviours, see Section~\ref{subsec:derive_asymp_normality}. Reviewing equation~\eqref{eq:expectation_on_hithat_hit0}, we can explicitly formulate $\left\{ h_t \left(0 \middle| \mathcal{F}_{t-1} \right) \right\}$ as follows:
\begin{equation}
h_t\left(0 \middle| \mathcal{F}_{t-1} \right) 
				=
				\mathbb{E}_{y_t,\widehat{\bm{\beta}}}\left[ \frac{ \text{Hit}_t(\widehat{\bm{\beta}}) - \text{Hit}_t(\bm{\beta}^{o})  }{ \nabla'f_t(\bm{\beta}^{o}) \left( \widehat{\bm{\beta}} - \bm{\beta}^{o} \right) } \middle| \mathcal{F}_{t-1} \right],
\label{eq:formulate_ht_role}
\end{equation}
which actually is a conditional expectation taken with respect to random variables $ y_t $ and $\widehat{\bm{\beta}} $. We use the subscript in $\mathbb{E}$ to clarify the expectation is taken with respect to specific random variable(s) hereafter.
Considering this role of $\left\{ h_t \left(0 \middle| \mathcal{F}_{t-1} \right) \right\}$ as well as equation~\eqref{eq:ht_role}, we are enlightened to use random bandwidth $\nabla'f_t(\widehat{\bm{\beta}}) \left( \bm{b}_i - \widehat{\bm{\beta}} \right)$ with $ \sqrt{T}\left( \bm{b}_i - \widehat{\bm{\beta}} \right) \dsim N(\bm{0},\bm{V_d})$ and $i = 1,2,\ldots, n$. We can set $\bm{V_d} = I_{(p+1)\times (p+1)}$ to start with. After sufficient $n$ times Monte Carlo simulating $\bm{b}_i - \widehat{\bm{\beta}} $ from $ N(\bm{0},\bm{V_d}) $, an estimator of $h_t \left(0 \middle| \mathcal{F}_{t-1} \right)$ can be achieved as follows: 
\begin{equation}
\begin{aligned}
\widehat{h_t}(0|\mathcal{F}_{t-1})
	& := 
	n^{-1}\Sum{n}{i=1} \frac{ \bm{1}\left\{ y_t \leq f_t(\widehat{\bm{\beta}})    \right\} - \bm{1}\left\{ y_t \leq f_t(\widehat{\bm{\beta}}) + \nabla'f_t(\widehat{\bm{\beta}}) \left( \bm{b}_i - \widehat{\bm{\beta}} \right)  \right\}  }{- \nabla'f_t(\widehat{\bm{\beta}}) \left( \bm{b}_i - \widehat{\bm{\beta}} \right) }.
\end{aligned}
\end{equation}
After achieving the above $\widehat{h_t}(0|\mathcal{F}_{t-1})$, we can estimate $\widehat{D}_T$ so as to update $\bm{V_d} = \widehat{D}_T^{-1}\widehat{A}_T\widehat{D}_T^{-1}$. Redo the simulation of $\{\bm{b}_i - \widehat{\bm{\beta}} \}_{i=1}^{n}$ with the updated $\bm{V_d}$. We can estimate $\widehat{h_t}(0|\mathcal{F}_{t-1})$ and $\widehat{D}_T$ again. This estimation repetition can mitigate the influence of an arbitrary chosen $\widehat{h_t}(0|\mathcal{F}_{t-1})$ in ARB.

Compared to the Powell Sandwich estimation~\eqref{eq:powell_sandwich} with $c_T$, our proposed method uses random bandwidth $\nabla'f_t(\widehat{\bm{\beta}}) \left( \bm{b}_i - \widehat{\bm{\beta}} \right)$ and Monte Carlo simulations such that it can adapt to time-varying conditional distributions of CAViaR DGPs by approaching to the role of $\left\{ h_t \left(0 \middle| \mathcal{F}_{t-1} \right) \right\}$ as in~\eqref{eq:ht_role} and in~\eqref{eq:formulate_ht_role}. 
The adaptive random bandwidth method can remarkably outperform the Powell Sandwich method in the applications on DGPs of time-varying conditional distributions, as shown in Table~\ref{tab:WT_ResABs_DGP_b0Nposy}. In theory, the adaptive random bandwidth method is valid as long as $\bm{b}_i - \widehat{\bm{\beta}}$ and $\bm{\beta}^{o} - \widehat{\bm{\beta}}$ have the same order of magnitude. We formally establish this adaptive random bandwidth method in Theorem~\ref{thm:est_ht_adaptive}.
\begin{theorem}[Adaptive Random Bandwidth Method]\label{thm:est_ht_adaptive}
{\ \\}
Assume the conditions and the asymptotic normality result in Theorem~\ref{thm:AN_CAViaR}. Choose an arbitrary positive definite symmetric matrix $\bm{V_d}$. Under the condition that 
\begin{equation}
\label{eq:bi_dist_condition}
\sqrt{T}\left( \bm{b}_i - \widehat{\bm{\beta}} \right) \overset{i.i.d.}{\sim}  N(\bm{0},\bm{V_d}), \qquad i = 1,\ldots,n, 
\end{equation}
and
\begin{equation*}
\biggl|\nabla'f_t(\widehat{\bm{\beta}}) \left( \bm{b}_i - \widehat{\bm{\beta}} \right) \biggr| \neq 0,
\end{equation*}
the adaptive random bandwidth estimator for $h_t(0|\mathcal{F}_{t-1})$ is formulated as follows:
\begin{equation}
\resizebox{1.08\textwidth}{!}{ $ 
\widehat{h_t}(0|\mathcal{F}_{t-1})
	 = 
	 \left\{
\begin{aligned}
	& 
	n^{-1}\Sum{n}{i=1} \frac{ \bm{1}\left\{ y_t \leq f_t(\widehat{\bm{\beta}}) + \nabla'f_t(\widehat{\bm{\beta}}) \left( \bm{b}_i - \widehat{\bm{\beta}} \right)  \right\}  - \bm{1}\left\{ y_t \leq f_t(\widehat{\bm{\beta}})    \right\} }{\nabla'f_t(\widehat{\bm{\beta}}) \left( \bm{b}_i - \widehat{\bm{\beta}} \right) },
	\quad & \text{when } y_t \neq f_t(\widehat{\bm{\beta}}),
	\\
	&
	0,
	\quad & \text{when } y_t = f_t(\widehat{\bm{\beta}}),	
\end{aligned}
\right. $}
\label{proposition_adaptive_random_bandwidth}
\end{equation}
such that
$$
\mathbb{E}_{y_t,\widehat{\bm{\beta}}}\biggl[ \widehat{h_t}(0|\mathcal{F}_{t-1}) \biggm|\mathcal{F}_{t-1}  \biggr]  \overset{p}{\longrightarrow} 
		  h_t(0|\mathcal{F}_{t-1})  
$$
as $n \rightarrow \infty$. \footnote{ We regard the least $(p+1)$ absolute residuals in $\{ | y_t - f_t(\widehat{\bm{\beta}})| \}_{t=1}^T$ as zeros. In fact, iterations of a simplex-based direct search method like the Nelder–Mead method for optimizing $(p+1)$ parameters terminates at the vertices of a simplex in the parameter space~\citep{lagarias1998convergence}. That is to say, the iterations in optimizing the $\tau$-th quantile regression objective function terminate with $(p+1)$ elements of $\{ ( \tau -\bm{1}\{ y_t - f_t(\bm{\beta} ) < 0 \} ) (y_t - f_t(\bm{\beta} ) ) \}$ solved to be zeros. Therefore, we set $\widehat{h_t}(0|\mathcal{F}_{t-1}) = 0$ at the least $(p+1)$ absolute residuals in $\{ | y_t - f_t(\widehat{\bm{\beta}})| \}_{t=1}^T$ in all the tests throughout this paper.}
\end{theorem}
\begin{proof}
See Appendix~\ref{sec:proofs}.
\end{proof}
\\

We separate the case of $y_t = f_t(\widehat{\bm{\beta}})$ from others to maintain the convergence of the ARB estimator due to $\lim_{x\rightarrow 0} \frac{1}{x} = \infty $. Zero given to $\widehat{h_t}(0|\mathcal{F}_{t-1})$ at $y_t = f_t(\widehat{\bm{\beta}})$ also enables the ARB estimator to approximate $h_t(0|\mathcal{F}_{t-1})$ from the left and from the right in half weights respectively in expectation, see the proof of Theorem~\ref{thm:est_ht_adaptive}.  
The convergence property of the partial sum in the sequence $\{\widehat{h_t}(0|\mathcal{F}_{t-1})\}$ by ARB is given in Corollary~\ref{prop:sequence_ht_adaptive}. 
\begin{corollary}
\label{prop:sequence_ht_adaptive}
Under the conditions of Theorem~\ref{thm:est_ht_adaptive}, the adaptive random bandwidth estimator $\{\widehat{h_t}(0|\mathcal{F}_{t-1})\}$ has the following property:
\begin{equation}
\label{eq:prop_average_htARB_covergence}
\frac{1}{T}\Sum{T}{t=1} \widehat{h_t}(0|\mathcal{F}_{t-1})
{\overset{m.s.}{\longrightarrow}}\, \frac{1}{T}\Sum{T}{t=1} h_t(0|\mathcal{F}_{t-1}),
\end{equation}
as $T, n \rightarrow \infty$. 
\end{corollary}
\begin{proof}
See Appendix~\ref{sec:proofs}.
\end{proof}
\\

It is clear that both $\widehat{\epsilon}_t := y_t - f_t(\widehat{\bm{\beta}})$ and $\nabla'f_t(\widehat{\bm{\beta}})$ are taken into account by ARB to approximate $h_t(0|\mathcal{F}_{t-1})$. In order to identify how $\widehat{\epsilon}_t$ and $\nabla'f_t(\widehat{\bm{\beta}})$ jointly shape $\widehat{h}_t(0|\mathcal{F}_{t-1})$, we would like to formulate $\widehat{h}_t(0|\mathcal{F}_{t-1})$ in Theorem~\ref{thm:est_ht_adaptive} into an analytic expression in terms of $\widehat{\epsilon}_t$ and $\nabla'f_t(\widehat{\bm{\beta}})$ so as to manifest the relationship. The analytic form of $\widehat{h_t}(0|\mathcal{F}_{t-1})$ by ARB described in Theorem~\ref{thm:est_ht_adaptive} is presented in Corollary~\ref{prop:analtic_ht_ARB}.    
\begin{corollary}
\label{prop:analtic_ht_ARB}
Under the conditions of Theorem~\ref{thm:est_ht_adaptive}, we can get the analytic form of $\widehat{h_t}(0|\mathcal{F}_{t-1})$ as follows:
\begin{equation}
\label{eq:analytic_ht_final}
\widehat{h_t}(0|\mathcal{F}_{t-1}) =
\left\{
\begin{aligned}
		& 
\frac{1}{2\delta_{\nabla_t}\,\sqrt{2\pi}} E_1\left( \frac{ \widehat{\epsilon}_t^2 }{2\delta_{\nabla_t}^2} \right),
	 & \quad \text{when } \widehat{\epsilon}_t \neq 0,
		\\
		& 
		0,
	& \quad \text{when } \widehat{\epsilon}_t = 0,
\end{aligned} 
\right.
\end{equation} 
where $\widehat{\epsilon}_t := y_t - f_t(\widehat{\bm{\beta}})$, $\delta_{\nabla}:=\sqrt{\frac{\nabla'f_t(\widehat{\bm{\beta}}) \bm{V_d}\nabla f_t(\widehat{\bm{\beta}})}{T}} = T^{-\frac{1}{2}}\norm{\nabla f_t}_2$, and $ E_1(s):= \int_{s}^{\infty} x^{-1}e^{-x} d\,x$ is a special integral known as the exponential integral or the incomplete gamma function $\Gamma(0,s)$. 
\end{corollary}
\begin{proof}
See Appendix~\ref{sec:proofs}.
\end{proof}
\\

\begin{figure}[hbtp]
\centering
\includegraphics[width = \textwidth, height = 8cm]{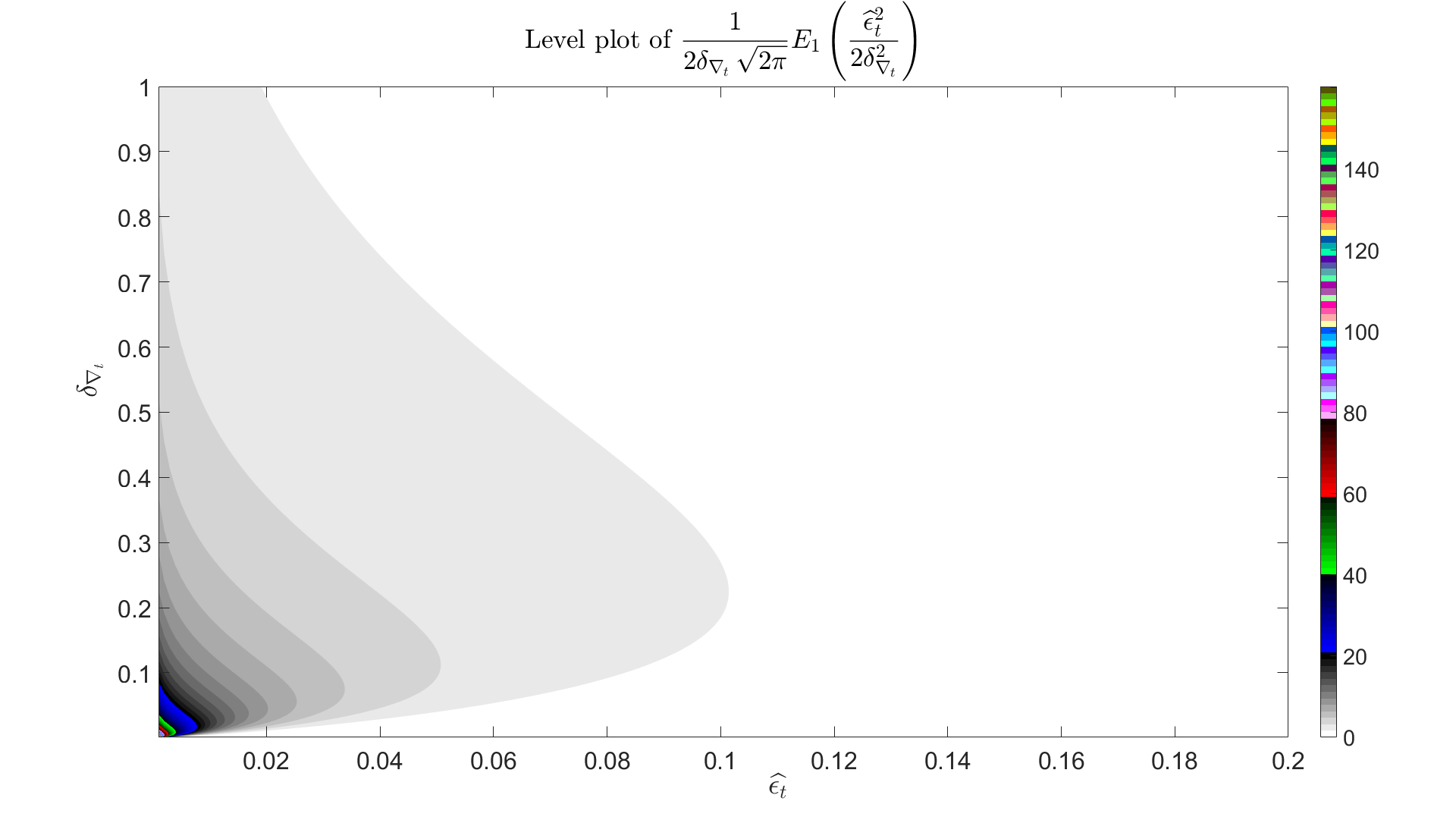}
\caption{Level plot for the analytic form of $\widehat{h_t}(0|\mathcal{F}_{t-1})$ by ARB in Corollary~\ref{prop:analtic_ht_ARB}.}
\label{fig:analytic_ht_epshat_delta}
\end{figure}

For visually checking the roles of $\widehat{\epsilon}_t $ and $ \delta_{\nabla_t} $ in the analytic $\widehat{h_t}(0|\mathcal{F}_{t-1})$ in Corollary~\ref{prop:analtic_ht_ARB}, we present a level plot of the analytic $\widehat{h_t}(0|\mathcal{F}_{t-1}) $ over $\widehat{\epsilon}_t $ and $ \delta_{\nabla_t} $ in Figure~\ref{fig:analytic_ht_epshat_delta} which uses colors to differentiate different ranges of  $\widehat{h_t}(0|\mathcal{F}_{t-1}) $. It is straightforward to get that the analytic $\widehat{h_t}(0|\mathcal{F}_{t-1})$ is decreasing in $|\widehat{\epsilon}_t |$ as also shown in Figure~\ref{fig:analytic_ht_epshat_delta}. However, $\delta_{\nabla_t} $, or say $T^{-\frac{1}{2}}\norm{\nabla f_t}_2$, can shift $\widehat{h_t}(0|\mathcal{F}_{t-1})$ by reflecting on how rare an $\widehat{\epsilon}_t $ is observed given the information set $\mathcal{F}_{t-1}$ and the model specification. That is how the information of $\delta_{\nabla_t} $ in ARB shapes $\widehat{h_t}(0|\mathcal{F}_{t-1})$ adaptively to time-varying conditional probability densities. 

The ARB estimator $\widehat{h_t}(0|\mathcal{F}_{t-1})$ via simulations in Theorem~\ref{thm:est_ht_adaptive} performs as robustly as the analytic ARB estimator $\widehat{h_t}(0|\mathcal{F}_{t-1})$ in Corollary~\ref{prop:analtic_ht_ARB}, as shown in Table~\ref{tab:WT_ResABs},~\ref{tab:WT_ResY} and~\ref{tab:WT_ResABs_DGP_b0Nposy}. The analytic way is faster than the simulation one. However, the ARB estimator via simulations is more intuitive and more flexible to adapt to a very different distribution for simulating $\{\bm{b}_i - \widehat{\bm{\beta}} \}_{i=1}^{n}$. 

$D_T$ need to be estimated consistently for inference tests on CAViaR models based on the asymptotic normality of the model parameter estimator. $\{\widehat{h_t}(0|\mathcal{F}_{t-1})\}$ by ARB facilitates our estimation on $D_T$ by just plugging in $ \widehat{\bm{\beta}}$ and
$\{\widehat{h_t}(0|\mathcal{F}_{t-1})\}$. The resulted estimator $\widehat{D}_T^{arb}$ has the consistency property presented in Theorem~\ref{thm:DThat_ARB}.
\begin{theorem}
\label{thm:DThat_ARB}
Under the conditions of Theorem~\ref{thm:est_ht_adaptive}, we can get that 
\begin{equation}
\widehat{D}_T^{arb}	 {\overset{p}{\longrightarrow}}\,  D_T,
\end{equation}
as $T \rightarrow \infty$ and $n \rightarrow \infty$, where $\widehat{D}_T^{arb}	 := 	 T^{-1} \Sum{T}{t=1} \widehat{h}_t\left(0 \middle| \mathcal{F}_{t-1} \right) \nabla'f_t( \widehat{\bm{\beta}}_{\tau} )  \nabla f_t(\widehat{\bm{\beta}}_{\tau} ) $ and $\widehat{h}_t\left(0 \middle| \mathcal{F}_{t-1} \right)$ is the adaptive random bandwidth estimator shown in~\eqref{proposition_adaptive_random_bandwidth}.
\end{theorem}
\begin{proof}
See Appendix~\ref{sec:proofs}.
\end{proof}
\\

The \textit{adaptive random bandwidth} (ARB) method is intuitive, robust and simple in practice, which can adapt to time-varying conditional distributions without a specific bandwidth or kernel function. A comparison of size performances of Wald tests using ARB with other competing methods are presented in Tables~\ref{tab:WT_ResABs}, ~\ref{tab:WT_ResY} and~\ref{tab:WT_ResABs_DGP_b0Nposy}. We also find that updating $\bm{V_d}$ improves the size performance with use of $\alpha$ levels in the interquartile range around but not much for $\alpha$ levels like $1\%, 5\%$. More test results are presented in Appendix~\ref{sec:appendix_tests} with changing sample size, quantile index and varying DGPs. The performance of ARB is robust. 
ARB can also be easily generalized to apply on multivariate quantile regressions, which is beyond the scope of this paper but in the interest of multivariate quantile regressions for future research. ARB also has the potential to achieve the second-order accuracy to Wald tests of nonlinear restrictions~\citep{phillips1988formulation,de1993corrections} in quantile regressions, which we would like to leave for future research.


\begin{table}
\centering
\caption{The size performances of the Wald test on the restricted model~\eqref{eq:CAViaR_RES_abs} to~\eqref{eq:CAViaR_Full_AS_Wald} with different estimation methods for  $\left\{ h_t(0|\mathcal{F}_{t-1}) \right\}$\;  ($\boldsymbol{\beta}_{0.5}^{R1} = [0,0.2,0.3,0.3], R= [0,0,1,-1], T=4000$)}
\label{tab:WT_ResABs}
\resizebox{\columnwidth}{!}{%
\begin{tabular}
[c]{l|ccccc}\hline\hline
Tests & size: $\alpha = 0.01$ & $\alpha = 0.05$ & $\alpha = 0.10$ & $\alpha = 0.20$ \\ \hline
Using $ \widehat{D}_T^{0} $	& 0.017 & 0.063 & 0.127  & 0.215
			\\
Using $ \widehat{D}_T^{h_0} $	& 0.016 & 0.066 & 0.131 & 0.215
			\\
Using $\widehat{D}_T^{arb}$ ($n=10^{4}$, $\bm{V_d} = \bm{I}_{4\times 4}$ with no update)     &  0.012 & 0.052 & 0.098 & 0.196 
	 		\\
Using $\widehat{D}_T^{arb}$ (analytic, $\bm{V_d} = \bm{I}_{4\times 4}$ with no update)	  &  0.012 & 0.052 & 0.102 & 0.198
	 		\\
Using $\widehat{D}_T^{arb}$ ($n=10^{4}$, 2 times updating $\bm{V_d}$)   	&  0.014 & 0.062 & 0.126 & 0.221
	 		\\
Using $\widehat{D}_T^{arb}$ (analytic, 2 times updating $\bm{V_d}$)   	& 0.014 & 0.061 & 0.125 & 0.219
			\\	 		
Using $\widehat{D}_T^{fd}$ ($\Delta\tau_T = \frac{10}{T}$)    	&  0.080	& 0.150 & 0.201 & 0.272
			\\			
Using $ \widehat{D}_T^{ker} $  &      0.017 & 0.069 & 0.129 & 0.223 
		 	\\ \hline
\end{tabular}
}
\end{table}

\begin{table}
\centering
\caption{The size performances of the Wald test on the restricted model~\eqref{eq:CAViaR_RES_Y} to~\eqref{eq:CAViaR_Full_AS_Wald} with different estimation methods for $\left\{ h_t(0|\mathcal{F}_{t-1}) \right\}$\;  ($\boldsymbol{\beta}_{0.5}^{R2} = [0,0.2,0.3,-0.3], R= [0,0,1,1],T=4000$)}
\label{tab:WT_ResY}
\resizebox{\columnwidth}{!}{%
\begin{tabular}
[c]{l|ccccc}\hline\hline
Tests & size: $\alpha = 0.01$ & $\alpha = 0.05$ & $\alpha = 0.10$ & $\alpha = 0.20$ \\ \hline
Using $ \widehat{D}_T^{0} $& 0.008 & 0.050 & 0.104  & 0.206

			\\
Using $ \widehat{D}_T^{h_0} $ & 0.007 & 0.050 & 0.105 & 0.207
			\\
Using $\widehat{D}_T^{arb}$ ($n=10^{4}$, $\bm{V_d} = \bm{I}_{4\times 4}$ with no update)   	&  0.01  & 0.046 & 0.084 & 0.168 
	 		\\
Using $\widehat{D}_T^{arb}$ (analytic, $\bm{V_d} = \bm{I}_{4\times 4}$ with no update)   	&  0.009 & 0.044 & 0.083 & 0.168
	 		\\	 		
Using $\widehat{D}_T^{arb}$ ($n=10^{4}$, 2 times updating $\bm{V_d}$)   	&   0.011 & 0.049 & 0.098 & 0.192
	 		\\
Using $\widehat{D}_T^{arb}$ (analytic, 2 times updating $\bm{V_d}$)   	&   0.01  & 0.048 & 0.097 & 0.19
	 		\\	 			 		
Using $\widehat{D}_T^{fd}$  ($\Delta\tau_T = \frac{10}{T}$)  &  0.049	& 0.104 & 0.153 & 0.229
			\\			
Using $ \widehat{D}_T^{ker} $  &   0.011 & 0.05  & 0.094 & 0.203 
		 	\\ \hline
\end{tabular}
}
\end{table}

\begin{table}
\centering
\caption{The size performances of the Wald test on the restricted model~\eqref{eq:timevarying_ht_CAViaR_DGP_b0Bposy_restricted} to~\eqref{eq:timevarying_ht_CAViaR_DGP_b0Bposy} with different estimation methods for  $\left\{ h_t(0|\mathcal{F}_{t-1}) \right\}$\;  ($\boldsymbol{\beta}_{0.5}^{R3} = [0,0.2,0.3,0.3], R= [0,0,1,-1], T=2000$)}
\label{tab:WT_ResABs_DGP_b0Nposy}
\resizebox{\columnwidth}{!}{%
\begin{tabular}
[c]{l|ccccc}\hline\hline
Tests & size: $\alpha = 0.01$ & $\alpha = 0.05$ & $\alpha = 0.10$ & $\alpha = 0.20$ \\ \hline
Using $\widehat{D}_T^{arb}$ ($n=10^{4}$, $\bm{V_d} = \bm{I}_{4\times 4}$ with no update)     & 0.024	& 0.052	 & 0.095	& 0.169
	 		\\
Using $\widehat{D}_T^{arb}$ (analytic, $\bm{V_d} = \bm{I}_{4\times 4}$ with no update)     & 0.023 &	0.054	& 0.093	& 0.168
	 		\\	 		
Using $\widehat{D}_T^{arb}$ ($n=10^{4}$, 2 times updating $\bm{V_d}$)	    & 0.021	& 0.055	& 0.095	& 0.188  		
	 		\\
Using $\widehat{D}_T^{arb}$ (analytic, 2 times updating $\bm{V_d}$)	    &  0.022	& 0.055	& 0.098 &	0.186		
	 		\\	 					
Using $ \widehat{D}_T^{ker} $  & 0.067	& 0.118	& 0.16	& 0.256 
		 	\\ \hline
\end{tabular}
}
\end{table}

%

\section{Empirical Results}
\label{sec:empirical_study}
We study four US stock prices which are the Dow Jones Composite Average (DJCA),
the NASDAQ 100 Index (NASDAQ100), the S\&P 500, and the Wilshire 5000 Total Market
Index (Will5000ind). We implement inference tests using the \textit{adaptive random bandwidth} method with $n=1000$ and $\bm{V_d} = \bm{I}_{(p+1)\times (p+1)}$ which is not updated in simulations in this section. Each stock price time series has 2448 daily prices,
ranging from 8th April 2010 to 30th December 2019. The price data were
converted to return rates by multiplying 100 with the difference of
the natural logarithm of the daily prices. The obtained return time series
of each stock contains 2447 observations which of the last 400 observations
are used for the out-of-sample testing after the first 2047 observations are
used to estimate the model.

The $5\%$ 1-day VaRs of a return time series are the opposite conditional $5\%$ 1-day quantiles of this time series. There are four different CAViaR models considered in this section to model the conditional quantiles of the stock return time series. The $5\%$ 1-day VaRs are estimated via the four different CAViaR specifications and the estimation results are shown in Table~\ref{tab:emp_AS_tau05},
~\ref{tab:emp_SAV_tau05}, ~\ref{tab:emp_IG_tau05} and~\ref%
{tab:emp_adapt_tau05} respectively. Each table contains the estimated parameters in a
specified model, the corresponding standard errors obtained by the \textit{%
adaptive random bandwidth} method with $n=1000$ and $\bm{V_d} = \bm{I}_{(p+1)\times (p+1)}$, the resulted two-sided
p-values on parameter significance, the optimized value of the quantile regression objective function (RQ), the percentage of times the VaR is exceeded, and the p-values of dynamic quantile (DQ) tests, both in-sample and out-of-sample. The model estimations, the in-sample DQ tests as well as the out-of-sample DQ tests in this empirical study are set up in the same way of Section~6 of~\cite{engle2004caviar}.  
\[
\left\{
\begin{tabular}{p{.8\textwidth}}
\begin{itemize}
\item Adaptive CAViaR:
\begin{equation}
f_t(\beta_1) = f_{t-1}(\beta_1) + \beta_{1} \left\{  \left[  1 + exp(G[y_{t-1} - f_{t-1}(\beta_1) ) ])  \right]^{-1} - \tau \right\},
\label{eq:adaptive_CAViaR}
\end{equation}
where $\tau$ is the quantile index of interest.
\item Symmetric absolute value CAViaR:
\begin{equation}
f_t(\boldsymbol{\beta}) = \beta_0 + \beta_1 f_{t-1}(\boldsymbol{\beta}) + \beta_{2}\, | y_{t-1} |.
\label{eq:SAV_CAViaR}
\end{equation}

\item Asymmetric slope CAViaR:
\begin{equation}
f_t(\boldsymbol{\beta}) = \beta_0 + \beta_1\, f_{t-1}(\boldsymbol{\beta}) + \beta_{2}\,\left( y_{t-1} \right)^{+} + \beta_{3}\, \left( y_{t-1} \right)^{-}.
\label{eq:AS_CAViaR}
\end{equation}

\item Indirect GARCH$(1,1)$:
\begin{equation}
f_t(\boldsymbol{\beta})  = -\sqrt{\beta_0 + \beta_1\, f_{t-1}(\boldsymbol{\beta})^2  + \beta_{2}\, y_{t-1}^2 }.
\label{eq:IGarch_CAViaR}
\end{equation}

\end{itemize}
\end{tabular}
\right.
\]
The above four CAViaR specifications have been defined as the adaptive
CAViaR, the symmetric absolute value CAViaR, the asymmetric slope CAViaR,
and the indirect GARCH$(1,1)$ respectively in the Section 3 of~\cite%
{engle2004caviar}. In the implementation of the adaptive model in this
emprical study, we follow \cite{engle2004caviar} and set $G = 10$. 
\begin{table}[htbp]
  \centering
  \caption{ The Asymmetric Slope Model ($\tau= 0.05$) }
  {\footnotesize
    \begin{tabular}{lrrrr}
    \toprule
    \toprule
    Stock Name & \multicolumn{1}{l}{DJCA } & \multicolumn{1}{l}{NASDAQ100} & \multicolumn{1}{l}{S\&P500} & \multicolumn{1}{l}{Will5000ind} \\
    \midrule
    \midrule
    $\hat{\beta}_0$ & -0.0538 & -0.1366 & -0.0772 & -0.0803 \\
    s.e.($\hat{\beta}_0$) & 0.0192 & 0.0324 & 0.0283 & 0.0259 \\
    p-value($\hat{\beta}_0$) & 0.0051* & 0.0000* & 0.0063* & 0.0019* \\
    \midrule
    $\hat{\beta}_1$ & 0.8913 & 0.8536 & 0.8651 & 0.8613 \\
    s.e.($\hat{\beta}_1$) & 0.0276 & 0.0356 & 0.0344 & 0.0344 \\
    p-value($\hat{\beta}_1$) & 0.0000* & 0.0000* & 0.0000* & 0.0000* \\
    \midrule
    $\hat{\beta}_2$ & -0.0175 & 0.0381 & 0.0264 & 0.0158 \\
    s.e.($\hat{\beta}_2$) & 0.0325 & 0.0717 & 0.0732 & 0.0831 \\
    p-value($\hat{\beta}_2$) & 0.5918 & 0.5950 & 0.7179 & 0.8487 \\
    \midrule
    $\hat{\beta}_3$ & -0.3069 & -0.3626 & -0.4249 & -0.4226 \\
    s.e.($\hat{\beta}_3$) & 0.0667 & 0.0673 & 0.1214 & 0.1153 \\
    p-value($\hat{\beta}_3$) & 0.0000* & 0.0000* & 0.0005* & 0.0002* \\
    \midrule
    RQ  & 205.1100 & 253.9000 & 215.3800 & 219.3200 \\
    Exceedance in-sample ( $\%$) & 5.0166 & 5.0639 & 5.0166 & 5.0166 \\
    Exceedance out-of-sample $\%$ & 4.7326 & 4.8746 & 4.5906 & 4.5433 \\
    DQ in-sample (p value) & 0.4306 & 0.5140 & 0.3094 & 0.4425 \\
    DQ out-of-sample (p value) & 1.0000 & 1.0000 & 1.0000 & 1.0000 \\
    \bottomrule
    \bottomrule
    \end{tabular}%
    }
  \label{tab:emp_AS_tau05}%
\end{table}%

\begin{table}[htbp]
  \centering
  \caption{ The Symmetric Absolute Value Model ($\tau= 0.05$) }
  {\footnotesize
    \begin{tabular}{lrrrr}
    \toprule
    \toprule
    Stock Name & \multicolumn{1}{l}{DJCA } & \multicolumn{1}{l}{NASDAQ100} & \multicolumn{1}{l}{S\&P500} & \multicolumn{1}{l}{Will5000ind} \\
    \midrule
    \midrule
    $\hat{\beta}_0$ & -0.0507 & -0.1310 & -0.0544 & -0.0521 \\
    s.e.($\hat{\beta}_0$) & 0.0405 & 0.0641 & 0.0430 & 0.0315 \\
    p-value($\hat{\beta}_0$) & 0.2103 & 0.0410* & 0.2064 & 0.0984 \\
    \midrule
    $\hat{\beta}_1$ & 0.8546 & 0.8127 & 0.8495 & 0.8676 \\
    s.e.($\hat{\beta}_1$) & 0.0418 & 0.0629 & 0.0544 & 0.0324 \\
    p-value($\hat{\beta}_1$) & 0.0000* & 0.0000* & 0.0000* & 0.0000* \\
    \midrule
    $\hat{\beta}_2$ & -0.2375 & -0.2492 & -0.2485 & -0.2161 \\
    s.e.($\hat{\beta}_2$) & 0.0266 & 0.0785 & 0.0775 & 0.0311 \\
    p-value($\hat{\beta}_2$) & 0.0000* & 0.0015* & 0.0013* & 0.0000* \\
    \midrule
    RQ    &  210.7300 & 263.0400 & 223.5300 & 227.3200 \\
    Exceedance in-sample ($\%$) & 5.0166 & 5.0166 & 5.0166 & 5.0166 \\
    Exceedance out-of-sample ($\%$) & 5.3952 & 5.2532 & 4.9219 & 4.9692 \\
    DQ in-sample (p value) & 0.2306 & 0.3548 & 0.0470* & 0.1537 \\
    DQ out-of-sample (p value) & 1.0000 & 1.0000 & 1.0000 & 1.0000 \\
    \bottomrule
    \bottomrule
    \end{tabular}%
    }
  \label{tab:emp_SAV_tau05}%
\end{table}%

\begin{table}[htbp]
  \centering
  \caption{The indirect GARCH$(1,1)$ ($\tau= 0.05$) }
  {\footnotesize
    \begin{tabular}{lrrrr}
    \toprule
    \toprule
    Stock Name & \multicolumn{1}{l}{DJCA } & \multicolumn{1}{l}{NASDAQ100} & \multicolumn{1}{l}{S\&P500} & \multicolumn{1}{l}{Will5000ind} \\
    \midrule
    \midrule
    $\hat{\beta}_0$ & 0.0651 & 0.2152 & 0.0878 & 0.0758 \\
    s.e.($\hat{\beta}_0$) &  0.0325 & 0.1069 & 0.0384 & 0.0414 \\
    p-value($\hat{\beta}_0$) & 0.0450* & 0.0442* & 0.0223* & 0.0670 \\
    \midrule
    $\hat{\beta}_1$ & 0.8741 & 0.7930 & 0.8290 & 0.8566 \\
    s.e.($\hat{\beta}_1$) & 0.0247 & 0.0444 & 0.0261 & 0.0258 \\
    p-value($\hat{\beta}_1$) & 0.0000* & 0.0000* & 0.0000* & 0.0000* \\
    \midrule
    $\hat{\beta}_2$ & 0.2551 & 0.3775 & 0.3638 & 0.2964 \\
    s.e.($\hat{\beta}_2$) & 0.2169 & 0.2031 & 0.2096 & 0.2041 \\
    p-value($\hat{\beta}_2$) & 0.2395 & 0.0631 & 0.0826 & 0.1465 \\
    \midrule
    RQ    &   209.4600 & 262.4600 & 222.1100 & 226.5200 \\
    Exceedance in-sample ($\%$) & 4.9692 & 5.0166 & 5.0639 & 5.0639 \\
    Exceedance out-of-sample ($\%$) & 5.3005 & 5.2059 & 4.6853 & 4.8273 \\
    DQ in-sample (p value) &  0.3678 & 0.4108 & 0.2887 & 0.4216 \\
    DQ out-of-sample (p value) & 1.0000 & 1.0000 & 1.0000 & 1.0000 \\
    \bottomrule
    \bottomrule
    \end{tabular}%
    }
  \label{tab:emp_IG_tau05}%
\end{table}%

\begin{table}[htbp]
  \centering
  \caption{ The Adaptive Model ($\tau= 0.05$) }
  {\footnotesize
    \begin{tabular}{lrrrr}
    \toprule
    \toprule
    Stock Name & \multicolumn{1}{l}{DJCA } & \multicolumn{1}{l}{NASDAQ100} & \multicolumn{1}{l}{S\&P500} & \multicolumn{1}{l}{Will5000ind} \\
    \midrule
    \midrule
    $\hat{\beta}_1$ &  -0.6980 & -0.7027 & -0.9827 & -1.5480 \\
    s.e.($\hat{\beta}_1$) & 0.0768 & 0.0760 & 0.0520 & 0.0014 \\
    p-value($\hat{\beta}_1$) & 0.0000* & 0.0000* & 0.0000* & 0.0000* \\
    \midrule
    RQ    &     213.4500 & 272.7100 & 226.9600 & 231.9700 \\
    Exceedance in-sample ( $\%$) & 4.4487 & 4.8746 & 4.6380 & 4.3067 \\
    Exceedance out-of-sample $\%$ & 4.7799 & 5.1585 & 4.8746 & 4.4960 \\
    DQ in-sample (p value) & 0.6518 & 0.9802 & 0.9545 & 0.2118 \\
    DQ out-of-sample (p value) & 1.0000 & 1.0000 & 1.0000 & 1.0000 \\
    \bottomrule
    \bottomrule
    \end{tabular}%
    }
  \label{tab:emp_adapt_tau05}%
\end{table}%

%
%

Comparing with the results in Section~6 of \cite{engle2004caviar}, we can see the standard errors obtained by the \textit{adaptive
random bandwidth} method is much smaller relatively to the size of estimated parameters. We use significance level 5\% to reject a parameter equal to zero as well as DQ tests. \lq\lq\, * \rq\rq\, denotes the rejections in Table~\ref{tab:emp_AS_tau05},~\ref{tab:emp_SAV_tau05},~\ref{tab:emp_IG_tau05} and~\ref{tab:emp_adapt_tau05}. Each of the four models shows almost the same rejection results for the stock return time series. Remarkably, it is observed that the coefficient $\beta_1$ of the VaR autoregressive term is highly significant from zero in all the four models for each stock return time series. This further supports the standpoint of CAViaR specifications, confirming that the
phenomenon of volatility clustering can be associated with the autoregressive VaR behaviour. 
The VaR exceedance in percentage indicates the realized risk level in applications. Dynamic quantile (DQ) tests based on the independence information regarding $\{ \text{Hit}_t\} $ are used to test model misspecification. We see a rejection in the in-sample DQ test on the symmetric absolute value model for the S\&P500 but the realized VaR exceedances (in-sample and out-of-sample) are much close to 5\% in Table~\ref{tab:emp_SAV_tau05}. So it can be complementary to judge CAViaR model specifications by looking at both VaR exceedances and inference tests like DQ tests.

In contract to the significance of $\beta_1$, the coefficient $\beta_2$ of $(y_{t-1})^{2}$ is insignificant in the indirect GARCH(1,1) model for all the stock return time series, see Table~\ref{tab:emp_IG_tau05}. And the coefficient $\beta_2$ of $(y_{t-1})^{+}$ is insignificant in the asymmetric slope model, see Table~\ref{tab:emp_AS_tau05}. Although the coefficient of $y_{t-1}$ is significant in the symmetric absolute model for all the stock return time series (see Table~\ref{tab:emp_SAV_tau05}), it is mainly due to the significant explanatory role of $(y_{t-1})^{-}$ based on the results of the asymmetric slope model which the symmetric absolute model is nested in. The significance results of $\beta_1$ in the adaptive model for each stock return time series suggest that the 5\% 1-day VaR can be associated with its 1-day lagged VaR violation which equals one if $y_{t-1}\leq f_{t-1}$ and zero otherwise. The significance results together implies that negative movements of a stock is significantly influential on its 5\% 1-day VaR in the next day.

In terms of the model goodness of fit, we look at the RQ results. The asymmetric slope model presents the lowest RQ result for each stock return time series among the four models despite that it has the most coefficients.

Overall, all the four stock return time series present the same strong associations with the lagged 5\% 1-day VaR in interpreting the present 5\% 1-day VaR. The asymmetric slope model and the adaptive CAViaR are satisfying for all the four stock returns in terms of data interpretation and model performance concerns. 


\section{Conclusions}
\label{sec:conclusion}
We found that the
inference test performance in CAViaR models is not robust and unsatisfying due to the estimation of the conditional probability densities of time series. We found that the existing density estimation methods cannot fully adapt to time-varying conditional probability densities of CAViaR time series. So in this paper we have developed a method called 
\textit{adaptive random bandwidth} which can robustly approximate the
time-varying conditional probability densities of CAViaR time series by
Monte Carlo simulations. This method not only avoids the haunting problem of choosing an optimal bandwidth but also ensures the reliability of CAViaR analysis based on the asymptotic normality of the model parameter estimator. In theory, our proposed method can be extended to general quantile regressions including multivariate cases easily and robustly. This method also has the potential to achieve the second-order accuracy to Wald tests of nonlinear restrictions~\citep{phillips1988formulation,de1993corrections} in quantile regressions.

\clearpage

\bibliographystyle{apalike}
\bibliography{biblio}

\newpage
\begin{appendices}
\section{Nonlinearity of parameters in CAViaR models}
\label{sec:nonlinearity_CAViaR}
Nonlinearity of parameters in CAViaR models differentiates CAViaR from linear quantile regressive models. In this appendix, we would like to illustrate the nonlinearity explicitly by showing the gradient, and the Hessian matrix of a CAViaR model.
\begin{equation}
f_t(\boldsymbol{\beta}_{\tau}) = \beta_0(\tau) + \beta_1 f_{t-1}(\boldsymbol{\beta}_{\tau}) + \beta_{2}\, \left( y_{t-1} \right)^{+} + \beta_{3}\, \left( y_{t-1} \right)^{-},
\label{eq:CAViaR_Full_AS}
\end{equation}
where $\tau \in (0,1)$, and the operators $(\cdot)^{+}$ and $(\cdot)^{-}$ are defined as $(x)^{+} = \max(x,0), (x)^{-} = -\min(x,0)$. This model can be rewritten by continuously substituting lagged conditional quantiles such as
\begin{equation}
\begin{aligned}
f_{t}(\boldsymbol{\beta}_{\tau}) 
						& = \beta_0(\tau) + \beta_1 f_{t-1}(\boldsymbol{\beta}_{\tau}) + \beta_{2}\, \left( y_{t-1} \right)^{+} + \beta_{3}\, \left( y_{t-1} \right)^{-}
						\\
						& = \beta_0(\tau) + \beta_1 \left( \beta_0(\tau) + \beta_1 f_{t-2}(\boldsymbol{\beta}_{\tau}) + \beta_{2}\, \left( y_{t-2} \right)^{+} + \beta_{3}\, \left( y_{t-2} \right)^{-} \right)
						\\
						& + \beta_{2}\, \left( y_{t-1} \right)^{+} + \beta_{3}\, \left( y_{t-1} \right)^{-}
						\\
						& = \frac{\beta_0(\tau)}{1 - \beta_1} +  \beta_{2}\,\Sum{\infty}{j=1} \beta_1^{j-1} \left( y_{t-j} \right)^{+} + \beta_{3}\,\Sum{\infty}{j=1} \beta_1^{j-1} \left( y_{t-j} \right)^{-},
\end{aligned}
\label{eq:explicit_CAViaR}
\end{equation}
where the last line comes from $ \abs{\beta_1}<1$. If $\beta_1 \neq 0$, \eqref{eq:explicit_CAViaR} reveals explicitly the nonlinear pattern of parameters in this CAViaR model. From this explicit form, we can further get the gradient and the Hessian matrix of the CAViaR model~\eqref{eq:CAViaR_Full_AS} to emphasize the roles of the parameters.

\subsection{$\nabla f_t$}
The gradient of $f_t(\boldsymbol{\beta}_{\tau})$ at a conditional quantile index $\tau\in (0,1)$ of interest can be derived as follows:
\begin{equation}
\begin{aligned}
& \nabla f_t(\boldsymbol{\beta}_{\tau})  = 
\begin{bmatrix}
\frac{\partial f_t(\boldsymbol{\beta}_{\tau})}{\partial \beta_0}
				\\
\frac{\partial f_t(\boldsymbol{\beta}_{\tau})}{\partial \beta_1}
				\\
				\frac{\partial f_t(\boldsymbol{\beta}_{\tau})}{\partial \beta_2}
				\\
				\frac{\partial f_t(\boldsymbol{\beta}_{\tau})}{\partial \beta_3}
 \end{bmatrix} 
 =
 \begin{bmatrix}
 1 + \beta_1\frac{\partial f_{t-1}(\boldsymbol{\beta}_{\tau})}{\partial \beta_0}
 			\\
f_{t-1}(\boldsymbol{\beta}_{\tau}) + \beta_1\frac{\partial f_{t-1}(\boldsymbol{\beta}_{\tau})}{\partial \beta_1}
 			\\ 
 \left( y_{t-1} \right)^{+} + \beta_1\frac{\partial f_{t-1}(\boldsymbol{\beta}_{\tau})}{\partial \beta_2}
 			\\
  \left( y_{t-1} \right)^{-} + \beta_1\frac{\partial f_{t-1}(\boldsymbol{\beta}_{\tau})}{\partial \beta_3}				
 \end{bmatrix}
 				\\
 				& = 
 \begin{bmatrix}
 1 + \beta_1\left( 1 + \beta_1\frac{\partial f_{t-1}(\boldsymbol{\beta}_{\tau})}{\partial \beta_0} \right)
 			\\
f_{t-1}(\boldsymbol{\beta}_{\tau}) + \beta_1\left( f_{t-2}(\boldsymbol{\beta}_{\tau}) + \beta_1\frac{\partial f_{t-2}(\boldsymbol{\beta}_{\tau})}{\partial \beta_1} \right)
 			\\ 
 \left( y_{t-1} \right)^{+} + \beta_1 \left(  \left( y_{t-2} \right)^{+} + \beta_1\frac{\partial f_{t-2}(\boldsymbol{\beta}_{\tau})}{\partial \beta_2} \right)
 			\\
 \left( y_{t-1} \right)^{-} + \beta_1 \left(  \left( y_{t-2} \right)^{-} + \beta_1\frac{\partial f_{t-2}(\boldsymbol{\beta}_{\tau})}{\partial \beta_3} \right)			
 \end{bmatrix} 	
 				 = 
 \begin{bmatrix}
 \frac{1}{1 - \beta_1}
 			\\
\Sum{\infty}{i=1} \beta_1^{i-1} f_{t-i}(\boldsymbol{\beta}_{\tau})
 			\\ 
\Sum{\infty}{i=1} \beta_1^{i-1}  \left( y_{t-i} \right)^{+}
 			\\
 \Sum{\infty}{i=1} \beta_1^{i-i} \left( y_{t-i} \right)^{-}	
 \end{bmatrix} .				
\end{aligned}
\label{eq:Dft}
\end{equation}
By knowing~\eqref{eq:explicit_CAViaR}, we substitute
\begin{equation*}
f_{t-i}(\boldsymbol{\beta}_{\tau}) 
						 = \frac{\beta_0(\tau)}{1 - \beta_1} +  \beta_{2}\,\Sum{\infty}{j=1} \beta_1^{j-1} \left( y_{t-i-j} \right)^{+} + \beta_{3}\,\Sum{\infty}{j=1} \beta_1^{j-1} \left( y_{t-i-j} \right)^{-},
\end{equation*}
into $\nabla f_t(\boldsymbol{\beta}_{\tau})$ in~\eqref{eq:Dft} and get
\begin{equation}
\begin{aligned}
\nabla f_t(\boldsymbol{\beta}_{\tau}) 
					& = 
 \begin{bmatrix}
 \frac{1}{1 - \beta_1}
 			\\
\Sum{\infty}{i=1} \beta_1^{i-1} \left(  \frac{\beta_0(\tau)}{1 - \beta_1} +  \beta_{2}\,\Sum{\infty}{j=1} \beta_1^{j-1} \left( y_{t-i-j} \right)^{+} + \beta_{3}\,\Sum{\infty}{j=1} \beta_1^{j-1} \left( y_{t-i-j} \right)^{-} \right)
 			\\ 
\Sum{\infty}{i=1} \beta_1^{i-1}  \left( y_{t-i} \right)^{+}
 			\\
 \Sum{\infty}{i=1} \beta_1^{i-i} \left( y_{t-i} \right)^{-}	
 \end{bmatrix} 
 					\\
 					& = 
 \begin{bmatrix}
 \frac{1}{1 - \beta_1}
 			\\
 \frac{\beta_0(\tau)}{(1 - \beta_1)^2} +  \beta_{2}\,\Sum{\infty}{h=2} (h-1) \beta_1^{h-2} \left( y_{t-h} \right)^{+} + \beta_{3}\,\Sum{\infty}{h=2} (h-1)\beta_1^{h-2} \left( y_{t-h} \right)^{-}
 			\\ 
\Sum{\infty}{i=1} \beta_1^{i-1}  \left( y_{t-i} \right)^{+}
 			\\
 \Sum{\infty}{i=1} \beta_1^{i-1} \left( y_{t-i} \right)^{-}	
 \end{bmatrix} . 
\end{aligned}	
\end{equation}
Now we can see the role of the parameters $\boldsymbol{\beta}_{\tau}$ explicitly. $\boldsymbol{\beta}_{\tau}$ shows up in all the elements of the gradient in a nonlinear form which makes it doubtless that the Hessian matrix does not fade out with $\boldsymbol{\beta}_{\tau}$ either. 

\subsection{Hessian matrix}
The second partial derivatives of $ f_t(\boldsymbol{\beta}_{\tau})$ exist as $\nabla f_t(\boldsymbol{\beta}_{\tau})$ does, which can be seen from the derivation of the Hessian matrix $\bm{H}(\boldsymbol{\beta}_{\tau})$ of $f_t(\boldsymbol{\beta}_{\tau})$ as follows:
\begin{equation}
\begin{aligned}
\bm{H} (\boldsymbol{\beta}_{\tau}) & = 
\begin{bmatrix}
\frac{\partial^2 f_t(\boldsymbol{\beta}_{\tau})}{\partial \beta_0^2}	
			&  
			\frac{\partial^2 f_t(\boldsymbol{\beta}_{\tau})}{\partial \beta_0 \partial \beta_1}
			&  
			\frac{\partial^2 f_t(\boldsymbol{\beta}_{\tau})}{\partial \beta_0 \partial \beta_2}
			&  
			\frac{\partial^2 f_t(\boldsymbol{\beta}_{\tau})}{\partial \beta_0 \partial \beta_3}
				\\
				\frac{\partial^2 f_t(\boldsymbol{\beta}_{\tau})}{\partial \beta_1 \partial \beta_0}
				&
				\frac{\partial^2 f_t(\boldsymbol{\beta}_{\tau})}{\partial \beta_1^2}
				&
				\frac{\partial^2 f_t(\boldsymbol{\beta}_{\tau})}{\partial \beta_1 \partial \beta_2}
				&
				\frac{\partial^2 f_t(\boldsymbol{\beta}_{\tau})}{\partial \beta_1 \partial \beta_3}
				\\
				\frac{\partial^2 f_t(\boldsymbol{\beta}_{\tau})}{\partial \beta_2 \partial \beta_0}	
			&  
			\frac{\partial^2 f_t(\boldsymbol{\beta}_{\tau})}{\partial \beta_2 \partial \beta_1}
			&  
			\frac{\partial^2 f_t(\boldsymbol{\beta}_{\tau})}{\partial \beta_2^2}
			&  
			\frac{\partial^2 f_t(\boldsymbol{\beta}_{\tau})}{\partial \beta_2 \partial \beta_3}
				\\
				\frac{\partial^2 f_t(\boldsymbol{\beta}_{\tau})}{\partial \beta_3 \partial \beta_0}
				&
				\frac{\partial^2 f_t(\boldsymbol{\beta}_{\tau})}{\partial \beta_3 \partial \beta_1}
				&
				\frac{\partial^2 f_t(\boldsymbol{\beta}_{\tau})}{\partial \beta_3 \partial \beta_2}
				&
				\frac{\partial^2 f_t(\boldsymbol{\beta}_{\tau})}{\partial \beta_3^2}
 \end{bmatrix} 	
 				\\
 				& = 
\resizebox{1.0\textwidth}{!}{
$
 \begin{bmatrix}
 0	&  (1 - \beta_1)^{-2}	& 0 & 0 
 			\\
 			\Sum{\infty}{i=1} \beta_1^{i-1} \frac{\partial f_{t-i}(\boldsymbol{\beta}_{\tau}) }{\partial \beta_0}
 			&
\Sum{\infty}{i=2} (i-1)\beta_1^{i-2} f_{t-i}(\boldsymbol{\beta}_{\tau}) + \Sum{\infty}{i=1} \beta_1^{i-1} \frac{\partial f_{t-i}(\boldsymbol{\beta}_{\tau}) }{\partial \beta_1}
			&
			\Sum{\infty}{i=1} \beta_1^{i-1} \frac{\partial f_{t-i}(\boldsymbol{\beta}_{\tau}) }{\partial \beta_2}	
			&
			\Sum{\infty}{i=1} \beta_1^{i-1} \frac{\partial f_{t-i}(\boldsymbol{\beta}_{\tau}) }{\partial \beta_3}		
 			\\ 
			0
			&
			\Sum{\infty}{i=2} (i-1)\beta_1^{i-2}  \left( y_{t-i} \right)^{+}
			&
			0
			&
			0
 			\\
 			0
 			&
			 \Sum{\infty}{i=2} (i-1)\beta_1^{i-2} \left( y_{t-i} \right)^{-}
			&
			0
			&
			0	
 \end{bmatrix} .
 $
}				
\end{aligned}
\label{eq:Hessian}
\end{equation}
Considering the rewritten form of, the gradient of, and the Hessian matrix of this CAViaR model, it might raise a caution of estimating those variables by using estimated parameters because the persistent appearance of the parameters can give a slow convergence rate. That is how in essence the nonlinearity of parameters in CAViaR models differentiates CAViaR from linear quantile regressive models.

\section{How to simulate CAViaR data generating processes}
\label{sec:simulate_CAViaR_DGPs}
Before estimating CAViaR models, we would like to provide a general way to simulate a time series $\{y_{t}\}$ of all conditional quantiles following a CAViaR specification. 
To generate such a CAViaR data generating process (DGP), it is required to get the information on the
parameter specification for every possible quantile so that the conditional
distribution of $\{y_{t}\}$ at each time can be constructed no matter
which quantile is realized. Indeed, when studying a data set, we might be interested in the 1\%-th, 5\%-th, 50\%-th
or 95\%-th conditional quantiles. For instance in the climate change literature,
extreme positive events are also of interest.

This requirement also applies when generating QAR DGPs. However,
simulating CAViaR models is more tedious than QAR simulations because the past
conditional distributions also need to be stored over time as they serve for
the CAViaR DGP simulation through the model VaR autoregressive terms each time. Let us illustrate the
simulation process through an example. First, we need to specify a CAViaR DGP at all quantiles for instance of ~\eqref{eq:CAViaR_AR} as follows:
\begin{equation*}
y_{t}=f_{t}(\boldsymbol{\beta }_{u_{t}})=\beta
_{0}(u_{t})+\sum\limits_{i=1}^{q}\beta _{i}(u_{t})\,f_{t-i}(\boldsymbol{%
\beta }_{u_{t}})+\sum\limits_{j=1}^{r}\beta _{q+j}(u_{t})\,y_{t-j},
\end{equation*}%
where $\boldsymbol{\beta _{u_{t}}}^{\prime }:=\left[ \beta _{0}(u_{t}),\beta
_{1}(u_{t}),\ldots ,\beta _{p}(u_{t})\right] $ with $p=q+r$, and $\left\{
u_{t}\right\} $ is i.i.d. in the standard uniform distribution (denoted as $\mathcal{U}(0,1)$). There is a monotonicity requirement on this model which is that $f_{t}(%
\boldsymbol{\beta }_{u_{t}})$ is monotonically increasing in $u_{t}$ so that
the $\tau $-th quantile ($\tau \in (0,1)$) of $%
y_{t}$ conditional on $\mathcal{F}_{t-1}$ can be expressed as $f_{t}(%
\boldsymbol{\beta _{\tau }})$. The additional step before simulating $%
\left\{ y_{t}\right\} _{t=1}^{T}$ is to specify the initial conditional distributions and the initial observations, i.e., $%
\left\{ f_{1-i}(\boldsymbol{\beta }_{\tau }),\tau \in (0,1),i=1,\ldots
,q\right\} $ and $\left\{ y_{1-j},j=1,\ldots ,r\right\} $. For example, we can take $%
f_{1-i}(\boldsymbol{\beta }_{\tau }) = F^{-1}_{N(0,1)}(\tau)$ for any $\tau \in (0,1),i=1,\ldots
,q $ and $y_{1-j}=0$ for $ j=1,\ldots ,r$, where $F^{-1}_{N(0,1)}(\tau)$ is denoted as the inverse function of the standard normal distribution.

With the above set-up, we can start the simulation by following the steps below. 


\begin{enumerate}
\item[Step 1:] Simulate a sequence of $\left\{ u_{t}\right\}
_{t=1}^{T}$ independently and identically distributed (i.i.d.) in $\mathcal{U}(0,1)$. $u_{t}$ indicates that $y_{t}$ is realized as its conditional $u_{t}$-th quantile.

\item[Step 2:] At time $t=1$, $y_{t}$ is realized as its $u_{t}$-th quantile which is equal to 
\begin{equation*}
f_{t}(\boldsymbol{\beta }_{u_{t}})=\beta
_{0}(u_{t})+\sum\limits_{i=1}^{q}\beta _{i}(u_{t})\,f_{t-i}(\boldsymbol{%
\beta }_{u_{t}})+\sum\limits_{j=1}^{r}\beta _{q+j}(u_{t})\,y_{t-j}.
\end{equation*}

\item[Step 3:] Store $\{ f_{t}(\boldsymbol{\beta }_{u_{t+k}}) \}_{k=1}^T$ by
\begin{equation*}
f_{t}(\boldsymbol{\beta }_{u_{t+k}})=\beta
_{0}(u_{t+k})+\sum\limits_{i=1}^{q}\beta _{i}(u_{t+k})\,f_{t-i}(\boldsymbol{%
\beta }_{u_{t+k}})+\sum\limits_{j=1}^{r}\beta _{q+j}(u_{t+k})\,y_{t-j}.
\end{equation*}
This step serves for generating $\{y_{t+k} \}_{k=1}^T$ later. For instance, $y_{t+k} = f_{t+k}(\boldsymbol{\beta }_{u_{t+k}})$ is generated via the information on $f_{t+k-i}(\boldsymbol{\beta }_{u_{t+k}}), i=1,\ldots,q$. Iteratively, it requires the conditional $u_{t+k}$-th quantiles of $\{ y_{t+k - i}\}_{i=1}^{t+k-1}$ to be stored for generating $y_{t+k}$.

\item[Step 4:] Repeat Step~2 and~3 for $t=2,3,\ldots ,T$ until we get $%
\left\{ y_{t}\right\} _{t=1}^{T}$. 

\item[Step 5:] In order to leave out the influence of the given initial values in this simulation, we have to delete the observations in the burn-in period. We delete the first 200 observations and keep the rest $\left\{ y_{t}\right\} _{t=201}^{T}$ as a suitable sample for studying the DGP~\eqref{eq:CAViaR_AR}.
\end{enumerate}

The above simulation procedure can be easily adapted to other CAViaR DGPs of which model equations of $f_{t}(\boldsymbol{\beta }_{\tau})$ can be substituted into Step~2 with observed values of any involved predetermined variables.

\section{Proofs}
\label{sec:proofs}
\subsection{Proof of Theorem~\ref{thm:est_ht_adaptive}}
\begin{proof}
\\
First, since expectation is a linear function, we can rewrite $\mathbb{E}_{y_t,\widehat{\bm{\beta}}}\left[ \widehat{h_t}(0|\mathcal{F}_{t-1})\middle|\mathcal{F}_{t-1}\right]$ as follows:
\begin{equation}
\begin{aligned}
		 & \mathbb{E}_{y_t,\widehat{\bm{\beta}}}\left[ \widehat{h_t}(0|\mathcal{F}_{t-1})\middle|\mathcal{F}_{t-1}\right] 
		\\
		=\, &
		n^{-1}\Sum{n}{i=1} \mathbb{E}_{y_t,\widehat{\bm{\beta}}} \left[  \frac{  \bm{1}\left\{ y_t \leq f_t(\widehat{\bm{\beta}}) + \nabla'f_t(\widehat{\bm{\beta}}) \left( \bm{b}_i - \widehat{\bm{\beta}} \right)  \right\}  - \bm{1}\left\{ y_t \leq f_t(\widehat{\bm{\beta}})    \right\} }{ \nabla'f_t(\widehat{\bm{\beta}}) \left( \bm{b}_i - \widehat{\bm{\beta}} \right) } \middle| \mathcal{F}_{t-1}, y_t \neq f_t(\widehat{\bm{\beta}}) \right]
		\\
		=\, &
		n^{-1}\Sum{n}{i=1} \mathbb{E}_{y_t,\widehat{\bm{\beta}}} \left[  \frac{  \bm{1}\left\{ 0< y_t - f_t(\widehat{\bm{\beta}}) \leq  \nabla'f_t(\widehat{\bm{\beta}}) \left( \bm{b}_i - \widehat{\bm{\beta}} \right)  \right\}  }{ \nabla'f_t(\widehat{\bm{\beta}}) \left( \bm{b}_i - \widehat{\bm{\beta}} \right) } \middle| \mathcal{F}_{t-1} \right]
		\\
		+\, &
		n^{-1}\Sum{n}{i=1} \mathbb{E}_{y_t,\widehat{\bm{\beta}}} \left[  \frac{  \bm{1}\left\{  \nabla'f_t(\widehat{\bm{\beta}}) \left( \bm{b}_i - \widehat{\bm{\beta}} \right)  < y_t - f_t(\widehat{\bm{\beta}})< 0  \right\} }{ -\nabla'f_t(\widehat{\bm{\beta}}) \left( \bm{b}_i - \widehat{\bm{\beta}} \right) } \middle| \mathcal{F}_{t-1}\right].
\end{aligned} 
\end{equation}
This equality holds when $n$ goes to infinity by applying the dominated convergence theorem as we regard the least $(p+1)$ absolute residuals in $\{ | y_t - f_t(\widehat{\bm{\beta}})| \}_{t=1}^T$ as zeros. Denote $\widehat{\epsilon}_t:= y_t - f_t(\widehat{\bm{\beta}})$. We rank $\{|\widehat{\epsilon}_t|\}_{t=1}^T$ from the smallest to largest into $\{|\widehat{\epsilon}|_{(1)},\cdot,|\widehat{\epsilon}|_{(T)}\}$. In fact, iterations of a simplex-based direct search method like the Nelder–Mead method for optimizing $(p+1)$ parameters terminates at the vertices of a simplex in the parameter space~\citep{lagarias1998convergence}. That is to say, the iterations in optimizing the $\tau$-th quantile regression objective function terminate with $(p+1)$ elements of $\{ ( \tau -\bm{1}\{ y_t - f_t(\bm{\beta} ) < 0 \} ) (y_t - f_t(\bm{\beta} ) ) \}$ solved to be zeros. Therefore, we set $\widehat{h_t}(0|\mathcal{F}_{t-1})$ at $\widehat{\epsilon}|_{(1)},\ldots,|\widehat{\epsilon}|_{(p+1)}$. And
\begin{equation}
\label{eq:bound_hthat}
| \widehat{h_t}(0|\mathcal{F}_{t-1}) | \leq \frac{1}{|\widehat{\epsilon}|_{(p+2)}} < \infty,
\end{equation}
where $|\widehat{\epsilon}|_{(p+2)} \neq 0 $ for a well-defined convex function minimization.

Since $\{ \bm{b}_i - \widehat{\bm{\beta}} \}_{i=1}^{n}$ is i.i.d in $N(\bm{0}, \bm{V_d})$ with restriction to
$
\nabla'f_t(\widehat{\bm{\beta}}) \left( \bm{b}_i - \widehat{\bm{\beta}} \right)  \neq 0,
$
we can get that for each $t\in\{1,\ldots,T\}$, 
$$
 \left\{ 
\mathbb{E}_{y_t,\widehat{\bm{\beta}}} \left[  \frac{  \bm{1}\left\{ y_t \leq f_t(\widehat{\bm{\beta}}) + \nabla'f_t(\widehat{\bm{\beta}}) \left( \bm{b}_i - \widehat{\bm{\beta}} \right)  \right\}  - \bm{1}\left\{ y_t \leq f_t(\widehat{\bm{\beta}})    \right\} }{ \nabla'f_t(\widehat{\bm{\beta}}) \left( \bm{b}_i - \widehat{\bm{\beta}} \right) } \middle|  \mathcal{F}_{t-1},  y_t \neq f_t(\widehat{\bm{\beta}}),  y_t \neq f_t(\widehat{\bm{\beta}})\right]
 \right\}_{i=1}^{n}
$$ 
is a sequence of independent random variables with finite second moments by the assumption of $\norm{\nabla'f_t(\widehat{\bm{\beta}})} \leq \bm{F_0}<\infty $ (see Assumption AN1(a) of~\cite{engle2004caviar} ). Then we can use Kolmogorov's strong Law of Large Number\citep[see e.g.][Corollary 3.9]{white2014asymptotic} and get that
\begin{equation}
\resizebox{1.08\textwidth}{!}{ $
\mathbb{E}_{y_t,\widehat{\bm{\beta}}}\left[ \widehat{h_t}(0|\mathcal{F}_{t-1})\middle|\mathcal{F}_{t-1}\right]
	\overset{a.s.}{\longrightarrow}
		\mathbb{E}_{y_t,\widehat{\bm{\beta}} , \bm{b}_i} \left[  \frac{  \bm{1}\left\{ y_t \leq f_t(\widehat{\bm{\beta}}) + \nabla'f_t(\widehat{\bm{\beta}}) \left( \bm{b}_i - \widehat{\bm{\beta}} \right)  \right\}  - \bm{1}\left\{ y_t \leq f_t(\widehat{\bm{\beta}})    \right\} }{ \nabla'f_t(\widehat{\bm{\beta}}) \left( \bm{b}_i - \widehat{\bm{\beta}} \right) } \middle| \mathcal{F}_{t-1}, y_t \neq f_t(\widehat{\bm{\beta}}) \right],
		$}
\label{eq:hthat_as_expectation}
\end{equation}
as $n\rightarrow\infty$ conditionally on $\mathcal{F}_{t-1}$. And we can further get that
\begin{equation}
\label{eq:expected_1_converge_ht}
 \begin{aligned}
& \mathbb{E}_{y_t,\widehat{\bm{\beta}} , \bm{b}_i} \left[  \frac{  \bm{1}\left\{ y_t \leq f_t(\widehat{\bm{\beta}}) + \nabla'f_t(\widehat{\bm{\beta}}) \left( \bm{b}_i - \widehat{\bm{\beta}} \right)  \right\}  - \bm{1}\left\{ y_t \leq f_t(\widehat{\bm{\beta}})    \right\} }{ \nabla'f_t(\widehat{\bm{\beta}}) \left( \bm{b}_i - \widehat{\bm{\beta}} \right) } \middle| \mathcal{F}_{t-1}, y_t \neq f_t(\widehat{\bm{\beta}}) \right]
		\\
&  = \mathbb{E}_{\widehat{\bm{\beta}}, \bm{b}_i} \left[ \mathbb{E}_{y_t} \left[  \frac{  \bm{1}\left\{ 0< y_t - f_t(\widehat{\bm{\beta}}) \leq  \nabla'f_t(\widehat{\bm{\beta}}) \left( \bm{b}_i - \widehat{\bm{\beta}} \right)  \right\}  }{ \nabla'f_t(\widehat{\bm{\beta}}) \left( \bm{b}_i - \widehat{\bm{\beta}} \right) } \middle| \mathcal{F}_{t-1}\right]  
\middle| \mathcal{F}_{t-1}\right]  
	\\
	& +
	\mathbb{E}_{\widehat{\bm{\beta}}, \bm{b}_i} \left[ \mathbb{E}_{y_t} \left[  
	\frac{  \bm{1}\left\{  \nabla'f_t(\widehat{\bm{\beta}}) \left( \bm{b}_i - \widehat{\bm{\beta}} \right)  < y_t - f_t(\widehat{\bm{\beta}})< 0  \right\} }{ -\nabla'f_t(\widehat{\bm{\beta}}) \left( \bm{b}_i - \widehat{\bm{\beta}} \right) } \middle| \mathcal{F}_{t-1}\right]  
\middle| \mathcal{F}_{t-1}\right] 
	\\
	& = 
	\mathbb{E}_{\widehat{\bm{\beta}}, \bm{b}_i} \left[ \frac{ F_t \left( f_t(\widehat{\bm{\beta}}) + \nabla'f_t(\widehat{\bm{\beta}}) \left( \bm{b}_i - \widehat{\bm{\beta}} \right)  \right)  - F_t\left( f_t(\widehat{\bm{\beta}})    \right) }{ \nabla'f_t(\widehat{\bm{\beta}}) \left( \bm{b}_i - \widehat{\bm{\beta}} \right) } \middle| \mathcal{F}_{t-1}, \nabla'f_t(\widehat{\bm{\beta}}) \left( \bm{b}_i - \widehat{\bm{\beta}} \right) > 0 \right] 
	\\
	& +
	\mathbb{E}_{\widehat{\bm{\beta}}, \bm{b}_i} \left[ \frac{ F_t\left( f_t(\widehat{\bm{\beta}})    \right) -  F_t \left( f_t(\widehat{\bm{\beta}}) + \nabla'f_t(\widehat{\bm{\beta}}) \left( \bm{b}_i - \widehat{\bm{\beta}} \right)  \right)   }{ -\nabla'f_t(\widehat{\bm{\beta}}) \left( \bm{b}_i - \widehat{\bm{\beta}} \right) } \middle| \mathcal{F}_{t-1}, \nabla'f_t(\widehat{\bm{\beta}}) \left( \bm{b}_i - \widehat{\bm{\beta}} \right) < 0 \right] 
	\\
	& = 
	\mathbb{E}_{\widehat{\bm{\beta}}, \bm{b}_i} \left[ \frac{ F_t'\left( f_t(\widehat{\bm{\beta}} )  \right)\nabla'f_t(\widehat{\bm{\beta}}) \left( \bm{b}_i - \widehat{\bm{\beta}} \right) + \mathcal{O}_p\left( \left( \bm{b}_i - \widehat{\bm{\beta}} \right)' \nabla f_t(\widehat{\bm{\beta}}) \nabla'f_t(\widehat{\bm{\beta}})  \left( \bm{b}_i - \widehat{\bm{\beta}} \right) \right) }{ \nabla'f_t(\widehat{\bm{\beta}}) \left( \bm{b}_i - \widehat{\bm{\beta}} \right) } \middle| \mathcal{F}_{t-1}\right]  
	\\
	& = 
	\mathbb{E}_{\widehat{\bm{\beta}}, \bm{b}_i} \biggl[ F_t'\left( f_t(\widehat{\bm{\beta}} )  \right) + \mathcal{O}_p\left( \nabla'f_t(\widehat{\bm{\beta}}) ( \bm{b}_i - \widehat{\bm{\beta}} ) \right) \biggm| \mathcal{F}_{t-1}\biggr]
	\\
	& = 
	\mathbb{E}_{\widehat{\bm{\beta}}, \bm{b}_i} \biggl[ F_t'\left( f_t(\widehat{\bm{\beta}} )  \right) \biggm| \mathcal{F}_{t-1}\biggr]
	\;  \overset{ T\rightarrow\infty}{\longrightarrow}  h_t(0|\mathcal{F}_{t-1}), 
\end{aligned}
\end{equation}
where the last two lines are obtained by Taylor's expansion for $F_t \left( f_t(\widehat{\bm{\beta}}) + \nabla'f_t(\widehat{\bm{\beta}}) \left( \bm{b}_i - \widehat{\bm{\beta}} \right)  \right)$ at $ f_t(\widehat{\bm{\beta}}) $ and by knowing $ \bm{b}_i - \widehat{\bm{\beta}} = o_p(1)$ and $\lim_{T\rightarrow\infty}\widehat{\bm{\beta}} =  \bm{\beta^{o}}$ with $ F_t'(f_t( \cdot) )$ being a continuous function (see AN1 and AN2 of~\citep{engle2004caviar}) respectively.

Therefore, we have $\mathbb{E}_{y_t}\left[ \widehat{h_t}(0|\mathcal{F}_{t-1}) \middle| \mathcal{F}_{t-1}) \right] - 
		 h_t(0|\mathcal{F}_{t-1})  = o_p(1)$ and conclude this proof.
\end{proof}

\subsection{Proof of Corollary~\ref{prop:sequence_ht_adaptive}}
\begin{proof}
From Theorem~\ref{thm:est_ht_adaptive}, we can obtain that 
\begin{equation}
\mathbb{E}\left[\frac{1}{T}\Sum{T}{t=1} \left( \widehat{h_t}(0|\mathcal{F}_{t-1})  -  h_t(0|\mathcal{F}_{t-1}) \right) \right]
=\, o_p(1) 
\end{equation}
because 
$
\mathbb{E}\left[ \widehat{h_t}(0|\mathcal{F}_{t-1})- h_t(0|\mathcal{F}_{t-1}) \right] 
		= o_p(1) 
$
when $n\rightarrow \infty$. And 
\begin{equation}
\label{eq:mean_square_difference_hthat}
\begin{aligned}
\frac{1}{T^2}\Sum{T}{t=1}\, & \mathbb{E}\left[ \left( \widehat{h_t}(0|\mathcal{F}_{t-1})  -  h_t(0|\mathcal{F}_{t-1}) \right)^2 \right] 
		\\
		& = \frac{1}{T^2}\Sum{T}{t=1} \mathbb{E}\left[  \widehat{h_t}^2(0|\mathcal{F}_{t-1})  -  h_t^2(0|\mathcal{F}_{t-1}) \right] + o_p(1).
\end{aligned}
\end{equation}
Denote $\widehat{h}_{t,i} := \frac{  \bm{1}\left\{ 0< y_t - f_t(\widehat{\bm{\beta}}) \leq  \nabla'f_t(\widehat{\bm{\beta}}) \left( \bm{b}_i - \widehat{\bm{\beta}} \right)  \right\}  }{ \nabla'f_t(\widehat{\bm{\beta}}) \left( \bm{b}_i - \widehat{\bm{\beta}} \right) }  + \frac{  \bm{1}\left\{  \nabla'f_t(\widehat{\bm{\beta}}) \left( \bm{b}_i - \widehat{\bm{\beta}} \right)  < y_t - f_t(\widehat{\bm{\beta}})< 0  \right\} }{ -\nabla'f_t(\widehat{\bm{\beta}}) \left( \bm{b}_i - \widehat{\bm{\beta}} \right) }, i=1,\ldots,n .$
We can derive that
\begin{equation}
\begin{aligned}
\mathbb{E}\left[ \widehat{h_t}^2(0|\mathcal{F}_{t-1})  \right]
		& =
		\mathbb{E}\left[
		\frac{2}{n^2} \Sum{n-1}{i=1}\Sum{n}{j=i+1}  \widehat{h}_{t,i}\widehat{h}_{t,j}
		\right]
		+
		\mathbb{E}\left[
		\frac{1}{n^2} \Sum{n}{i=1}  \widehat{h}_{t,i}^2
		\right]
		\\
		& =
		\frac{2}{n^2} \Sum{n-1}{i=1}\Sum{n}{j=i+1} \mathbb{E}\left[ \widehat{h}_{t,j} \widehat{h}_{t,i} \right]
		+
		\mathbb{E}\left[
		\frac{1}{n^2} \Sum{n}{i=1}  \widehat{h}_{t,i}^2
		 \right]
		\\
		& =
		\frac{2}{n^2}\frac{n(n-1)}{2} \mathbb{E}\left[
		F_t^{'2}\left( f_t(\widehat{\bm{\beta}} )  \right) 
		+
		 \mathcal{O}_p\left( \nabla'f_t(\widehat{\bm{\beta}}) ( \bm{b}_i - \widehat{\bm{\beta}} ) \right)		
		\right] + \mathbb{E}\left[
		\frac{1}{n^2} \Sum{n}{i=1}  \widehat{h}_{t,i}^2
		 \right]
		\\
		& \overset{ T\rightarrow\infty}{\longrightarrow}
		\mathbb{E}\left[ h_t^2(0|\mathcal{F}_{t-1})\right]
		+ 
		\frac{1}{n} \mathbb{E}\left[ h_t^2(0|\mathcal{F}_{t-1}) \right]
		+ o_p(1)
		+ 
		\frac{1}{n^2} \Sum{n}{i=1} \mathbb{E}\left[
		 \widehat{h}_{t,i}^2
		 \right]
		\\
		& =
		\mathbb{E}\left[ h_t^2(0|\mathcal{F}_{t-1})\right]
		+
		\mathcal{O}_p(\frac{1}{n}),
\end{aligned}
\label{eq:hthat_square}
\end{equation}
which follows the reasoning of~\eqref{eq:expected_1_converge_ht}, and herein the last line is obtained by knowing $\{ h_t(0|\mathcal{F}_t)\}$ and $\{\widehat{h}_t(0|\mathcal{F}_t)\}$ is uniformly bounded by a finite constant according to Assumption AN2 of~\cite{engle2004caviar} and~\eqref{eq:bound_hthat} respectively. 
Now substitute~\eqref{eq:hthat_square} back to~\eqref{eq:mean_square_difference_hthat} and get that
\begin{equation}
\begin{aligned}
\frac{1}{T^2}\Sum{T}{t=1}\, &  \mathbb{E}\left[ \left( \widehat{h_t}(0|\mathcal{F}_{t-1})  -  h_t(0|\mathcal{F}_{t-1}) \right)^2 \right] 
		\\		
		& =  \frac{1}{T^2}\Sum{T}{t=1} \mathbb{E}\left[  h_t^2(0|\mathcal{F}_{t-1})  -  h_t^2(0|\mathcal{F}_{t-1}) \right] 
		+
		\mathcal{O}_p(\frac{1}{T\,n})
		\\
		& = \mathcal{O}_p(\frac{1}{T\,n}),
\end{aligned}	
\end{equation}
which leads to 
\begin{equation}
\frac{1}{T^2}\Sum{T}{t=1}\, \mathbb{E}\left[ \left( \widehat{h_t}(0|\mathcal{F}_{t-1})  -  h_t(0|\mathcal{F}_{t-1}) \right)^2 \right] \overset{p}{\rightarrow} 0
\end{equation}
when $T,n\rightarrow\infty$.
Therefore, we obtain the mean square convergence~\eqref{eq:prop_average_htARB_covergence} for the mean of the adaptive random bandwidth estimator sequence $\{\widehat{h_t}(0|\mathcal{F}_{t-1})\}$. 

\end{proof}

\subsection{Proof of Corollary~\ref{prop:analtic_ht_ARB}}
\begin{proof}
From the condition~\eqref{eq:bi_dist_condition}, we can know that
$$
\nabla'f_t(\widehat{\bm{\beta}})\left( \bm{b}_i - \widehat{\bm{\beta}} \right) \overset{i.i.d.}{\sim}  N(\bm{0}, \frac{1}{T}\,\nabla'f_t(\widehat{\bm{\beta}}) \,\bm{V_d}\,\nabla f_t(\widehat{\bm{\beta}})), \qquad i = 1,\ldots,n,
$$
and
\begin{equation*}
\biggl|\nabla'f_t(\widehat{\bm{\beta}}) \left( \bm{b}_i - \widehat{\bm{\beta}} \right) \biggr| \neq 0.
\end{equation*}
Denote the probability distribution function of $\nabla'f_t(\widehat{\bm{\beta}})\left( \bm{b}_i - \widehat{\bm{\beta}} \right)$ as $F_{\nabla}(\cdot)$, $\delta_{\nabla}:=\sqrt{\frac{\nabla'f_t(\widehat{\bm{\beta}}) \,\bm{V_d}\, \nabla f_t(\widehat{\bm{\beta}})}{T}}$ and $\widehat{\epsilon}_t:= y_t - f_t(\widehat{\bm{\beta}})$.

From~\eqref{eq:hthat_as_expectation} in the proof of Theorem~\ref{thm:est_ht_adaptive}, we know that
\begin{equation}
\begin{aligned}
 		 \widehat{h_t}(0|\mathcal{F}_{t-1})
		& \overset{n\rightarrow\infty}{\longrightarrow}
		\mathbb{E}_{\bm{b}_i} \left[  \frac{  \bm{1}\left\{ y_t \leq f_t(\widehat{\bm{\beta}}) + \nabla'f_t(\widehat{\bm{\beta}}) \left( \bm{b}_i - \widehat{\bm{\beta}} \right)  \right\}  - \bm{1}\left\{ y_t \leq f_t(\widehat{\bm{\beta}})    \right\} }{ \nabla'f_t(\widehat{\bm{\beta}}) \left( \bm{b}_i - \widehat{\bm{\beta}} \right) } \middle| \mathcal{F}_{t-1}, y_t \neq f_t(\widehat{\bm{\beta}})\right]
		\\
		& = \int_{\mathbb{R}\setminus[-|\widehat{\epsilon}_t|,|\widehat{\epsilon}_t|)} \frac{  \bm{1}\left\{ y_t \leq f_t(\widehat{\bm{\beta}}) + x  \right\}  - \bm{1}\left\{ y_t \leq f_t(\widehat{\bm{\beta}})    \right\} }{ x } d\,F_v(x).
\end{aligned} 
\label{eq:hthat_expectation_int0}
\end{equation}
We can further rewrite \eqref{eq:hthat_expectation_int0} based on two cases in $\widehat{\epsilon}_t$, namely $\widehat{\epsilon}_t > 0$, $\widehat{\epsilon}_t < 0$ since $\widehat{h_t}(0|\mathcal{F}_{t-1})$ is set to be zero in ARB when $\widehat{\epsilon}_t = 0$.

When $\widehat{\epsilon}_t > 0$, we get 
\begin{equation}
\label{eq:hthat_expectation_int1}
\begin{aligned}
\widehat{h_t}(0|\mathcal{F}_{t-1})
		& = \int_{\mathbb{R}\setminus[-|\widehat{\epsilon}_t|,|\widehat{\epsilon}_t|)}  \frac{  \bm{1}\left\{ y_t \leq f_t(\widehat{\bm{\beta}}) + x  \right\}  - \bm{1}\left\{ y_t \leq f_t(\widehat{\bm{\beta}})    \right\} }{ x } d\,F_v(x)
		\\
		& =\int_{\mathbb{R}\setminus[-|\widehat{\epsilon}_t|,|\widehat{\epsilon}_t|)} \frac{  \bm{1}\left\{ \widehat{\epsilon}_t \leq x  \right\}  }{ x } d\,F_v(x)
		\\
		& = \int_{ \widehat{\epsilon}_t}^{\infty} \frac{  1 }{ x } \frac{1}{\delta_{\nabla_t}\,\sqrt{2\pi}}\,e^{-\frac{x^2}{2\delta^2_{\nabla}}} d\,x.
\end{aligned} 
\end{equation}  
Substitute $u:=\frac{x^2}{2\delta_{\nabla_t}^2}$ into~\eqref{eq:hthat_expectation_int1} and get 
\begin{equation}
\label{eq:hthat_expectation_int2}
\begin{aligned}
\widehat{h_t}(0|\mathcal{F}_{t-1})
		& = \frac{1}{\delta_{\nabla_t}\,\sqrt{2\pi}} \frac{1}{2}\int_{\frac{\widehat{\epsilon}_t^2}{2\delta_{\nabla_t}^2}} ^{\infty} \frac{  e^{-u} }{ u } \, d\,u
		\\
		& =  \frac{1}{2\delta_{\nabla_t}\,\sqrt{2\pi}} E_1\left( \frac{\widehat{\epsilon}_t^2}{2\delta_{\nabla_t}^2} \right),
\end{aligned} 
\end{equation} 
where $ E_1(s):= \int_{s}^{\infty} x^{-1}e^{-x} d\,x$ is a special integral known as the exponential integral or the incomplete gamma function $\Gamma(0,s)$.

Analogously, when $\widehat{\epsilon}_t < 0$, we can also get
\begin{equation}
\label{eq:hthat_expectation_int}
\widehat{h_t}(0|\mathcal{F}_{t-1})
		 =  \frac{1}{2\delta_{\nabla_t}\,\sqrt{2\pi}} E_1\left( \frac{\widehat{\epsilon}_t^2}{2\delta_{\nabla_t}^2} \right).
\end{equation}

Therefore, we conclude this proof.
\end{proof}

\subsection{Proof of Theorem~\ref{thm:DThat_ARB}}
\begin{proof}
Denote that
\begin{equation}
\bar{D}_T := T^{-1} \Sum{T}{t=1} h_t\left(0 \middle| \mathcal{F}_{t-1} \right) \nabla'f_t( \widehat{\bm{\beta}}_{\tau} )  \nabla f_t(\widehat{\bm{\beta}}_{\tau} ).
\end{equation}
Note that
\begin{equation}
\widehat{D}_T^{arb} - D_T = \widehat{D}_T^{arb} - \bar{D}_T + \bar{D}_T - D_T.
\end{equation}
It is straightforward to get that
\begin{equation}
\widehat{D}_T^{arb} - \bar{D}_T  = o_p \left( 1 \right),
\end{equation}
since we know that
$$
 \frac{1}{T}\Sum{T}{t=1} \widehat{h_t}(0|\mathcal{F}_{t-1})
- \frac{1}{T}\Sum{T}{t=1} h_t(0|\mathcal{F}_{t-1}) = o_p(1)
$$
from $\frac{1}{T}\Sum{T}{t=1} \widehat{h_t}(0|\mathcal{F}_{t-1})
{\overset{m.s.}{\longrightarrow}}\, \frac{1}{T}\Sum{T}{t=1} h_t(0|\mathcal{F}_{t-1})$ given in Corollary~\ref{prop:sequence_ht_adaptive} with $\{ \nabla f_t(\bm{\beta})\nabla'f_t(\bm{\beta}) \}$ being uniformly bounded in $\mathbb{R}^{p+1}$ by Assumption AN1 of~\cite{engle2004caviar}.
And 
\begin{equation}
\bar{D}_T - D_T = o_p \left( 1 \right),
\end{equation}
since that $\widehat{A}_T -  A_T{\overset{p}{\longrightarrow}}\, 0 $ which has been proved in Theorem~3 of~\cite{engle2004caviar} and $\{ h_t(\cdot |\mathcal{F}_{t-1})\}$ is uniformly bounded by a finite constant according to Assumption AN2 of~\cite{engle2004caviar}.

Therefore, we have that $\widehat{D}_T^{arb} - D_T = o_p \left(1\right)$ and conclude this proof.
\end{proof}

\clearpage

\section{Extra figures}
\label{sec:appendix_figs}
\vspace{-0.5cm}
\begin{figure}[hbtp]
\centering
\hspace*{-1.5cm}\includegraphics[width = 1.2\textwidth, height =
0.72\textheight]{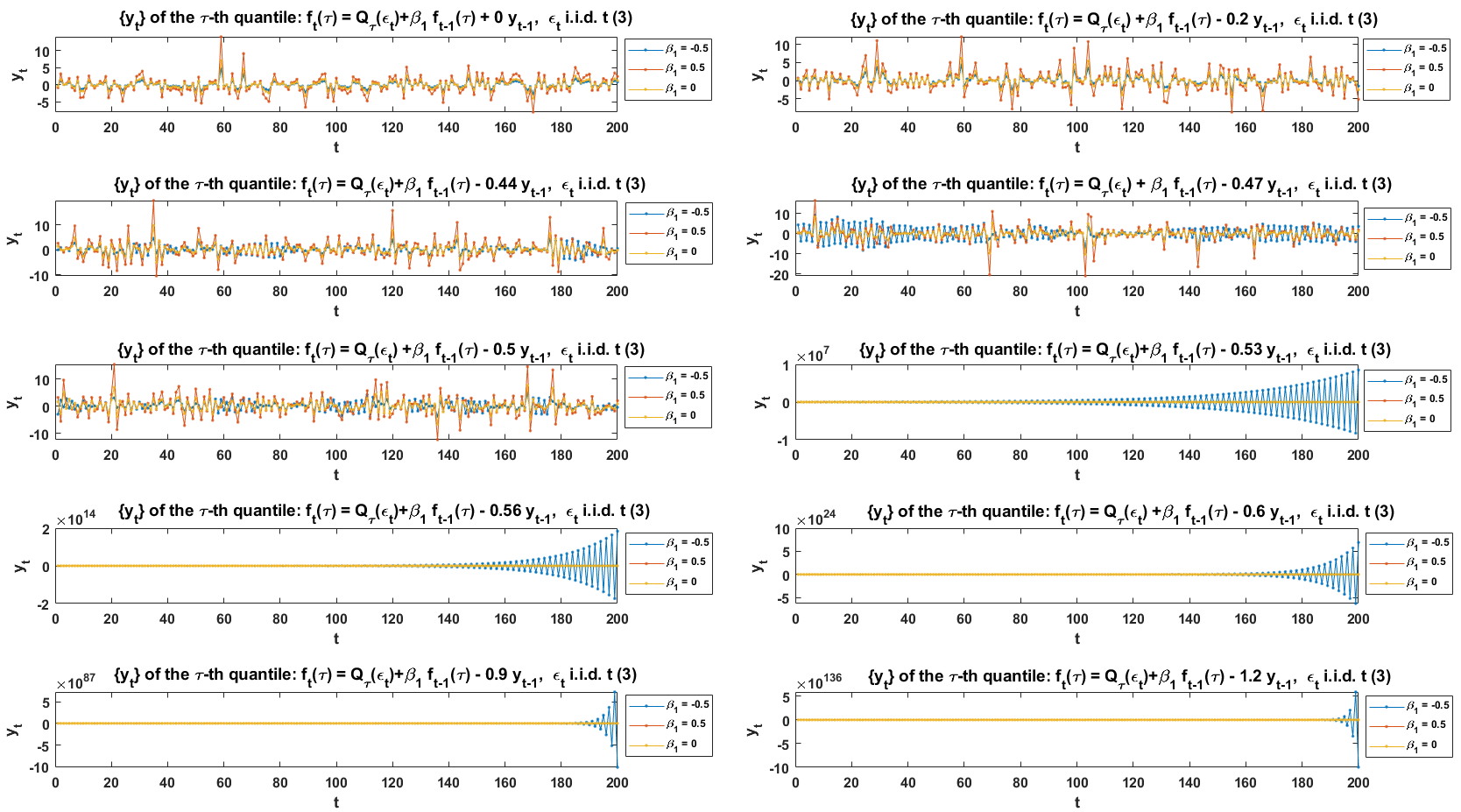}
\caption{Time series plots of CAViaR DGP samples for illustration}
\label{fig:ts_DGPs_stability_check1}
\end{figure}

\begin{figure}[hbtp]
\centering
\hspace*{-1.5cm}\includegraphics[width = 1.2\textwidth, height = 0.9
\textheight]{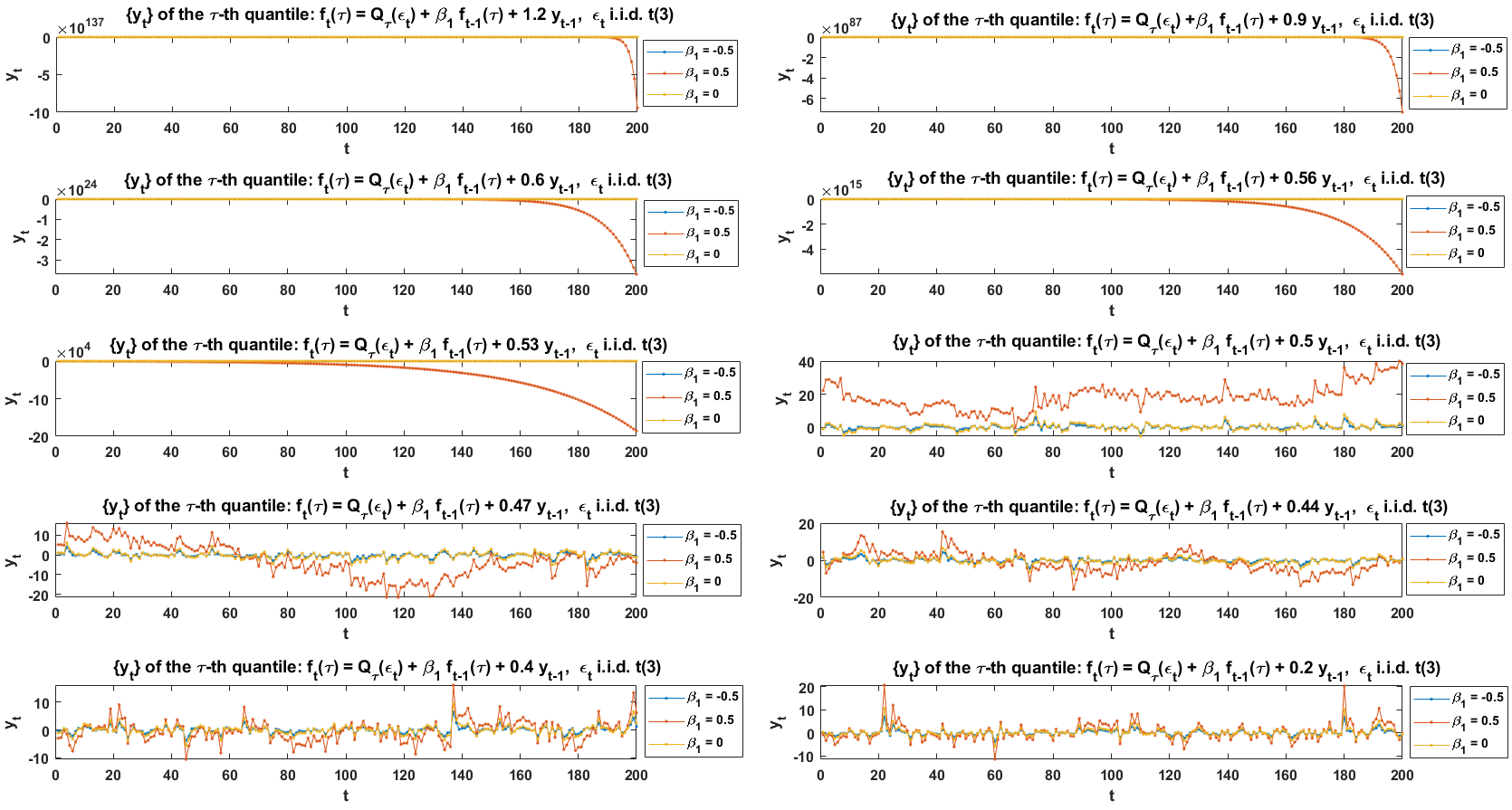}
\caption{Time series plots of CAViaR DGP samples for illustration}
\label{fig:ts_DGPs_stability_check2}
\end{figure}

\clearpage

\section{Extra test results}
\label{sec:appendix_tests}

\begin{itemize}
\item
Simulate 1000 samples from the following DGP:
\begin{equation}
\begin{aligned}
y_t = f_t(\bm{\beta}_{u_t}^{R4}) 
								& = \beta_0^{R4}(u_t) + \beta_1^{R4}(u_t) f_{t-1}(\boldsymbol{\beta}_{\tau}^{R4}) + \beta_{2}^{R4}(u_t)\, | y_{t-1} |
								\\
								& = \beta_0^{R4}(u_t) + \beta_1^{R4}(u_t) f_{t-1}(\boldsymbol{\beta}_{u_t}^{R4}) + \beta_{2}^{R4}(u_t)\, \left( y_{t-1} \right)^{+} + \beta_{2}^{R4}(u_t)\, \left( y_{t-1} \right)^{-},
\end{aligned}
\label{eq:timevarying_ht_CAViaR_DGP_3regimeb0_b2BETA}
\end{equation}
where $\left\{ u_t \right\}\overset{i.i.d.}{\sim} \mathcal{U}(0,1)$ and the underlying parameters change over $u_t$ as follows:
\begin{equation}
\left\{
\begin{aligned}
\beta_0^{R4}(u_t) & =
		\left\{
		\begin{aligned}
		3\,F^{-1}_{N(0,1)} \,, \qquad\, 0<u_t\leq 0.4;
		\\
		F^{-1}_{N(0,1)} \,, \qquad\, 0.4<u_t\leq 0.6;
		\\
		2\,F^{-1}_{N(0,1)} \,, \qquad\, 0.6<u_t < 1,
		\end{aligned}
		\right.
		\\
\beta_1^{R4}(u_t) & =0.2, \qquad\, 0 <u_t < 1,
		\\
\beta_2^{R4}(u_t) & =0.3, \qquad\, 0 <u_t < 1,	
\end{aligned}
\right.
\end{equation}
where $F^{-1}_{N(0,1)}(\cdot)$ is the inverse standard normal probability distribution function.
Conditional $5\%$-th, $30\%$-th, $50\%$-th quantiles are estimated for each of the total $1000$ simulated samples of sample size $T$ by regressing the sample onto the full model~\eqref{eq:CAViaR_Full_AS_Wald}. 
The results of the Wald test using the adaptive random bandwidth method and the kernel method~\eqref{eq:parzon_window} are listed in Table~\ref{tab:WT_ResABs_tau_T_3regime_b2beta} in which each estimated size is obtained by the percentage rejection rate among the 1000 samples of sample size $T$.

\begin{table}[h]
\centering
\caption{The size performances of the Wald test on the restricted model~\eqref{eq:CAViaR_RES_abs} to~\eqref{eq:CAViaR_Full_AS_Wald} \;  ($\bm{\beta}_{u_t}^{R4} = [F^{-1}_{N(0,1)}(u_t), 0.2, 0.3]'$, $R= [0,0,1,-1]$)}
\label{tab:WT_ResABs_tau_T_3regime_b2beta}
\resizebox{\columnwidth}{!}{%
\begin{tabular}
[c]{l|l|ccccc}\hline\hline
quantile index $\tau$ \& sample size T & methods & size: $\alpha = 0.01$ & $  \alpha = 0.05$ & $\alpha = 0.10$ & $\alpha = 0.20$ \\ \hline
\multirow{3}{*}{$\tau = 0.05, T=5000 $} & $\widehat{D}_T^{arb}$ ($n=10^{4}$,    &       0.016 & 0.054 & 0.091 & 0.179 	
	 		\\
	 		&  2 times updating $\bm{V_d}$)    &  	&	&	 	&
	 \\			
& $ \widehat{D}_T^{ker} $  & 	0.026 & 0.066 & 0.126 & 0.21
		 	\\ \hline
\multirow{3}{*}{$\tau = 0.05, T=2000 $} & $\widehat{D}_T^{arb}$ ($n=10^{4}$,    &     0.024 & 0.08  & 0.134 & 0.228 
	 		\\
	 		&  2 times updating $\bm{V_d}$)    &  	&	&	 	&
	 \\			
& $ \widehat{D}_T^{ker} $  & 0.036 & 0.107 & 0.176 & 0.288
			\\ \hline
\multirow{3}{*}{$\tau = 0.3, T=5000 $} & $\widehat{D}_T^{arb}$ ($n=10^{4}$,    &   	0.01  & 0.045 & 0.085 & 0.168	
	 		\\
	 		&  2 times updating $\bm{V_d}$)    &  & & &
	 \\			
& $ \widehat{D}_T^{ker} $  &  0.011 & 0.053 & 0.095 & 0.182	
		 	\\ 
		 	\hline  
\multirow{3}{*}{$\tau = 0.3, T=2000 $} & $\widehat{D}_T^{arb}$ ($n=10^{4}$,    &    0.015 & 0.049 & 0.085 & 0.192	
	 		\\
	 		&  2 times updating $\bm{V_d}$)    &  	&	&	 	&
	 \\			
& $ \widehat{D}_T^{ker} $  & 	0.009 & 0.036 & 0.091 & 0.197
		 	\\ 	\hline  
\multirow{3}{*}{$\tau = 0.5, T=5000 $} & $\widehat{D}_T^{arb}$ ($n=10^{4}$,    &   0.014 & 0.056 & 0.087 & 0.18 		
	 		\\
	 		&  2 times updating $\bm{V_d}$)    &  	&	&	 	&
	 \\		
& $ \widehat{D}_T^{ker} $  & 	  0     & 0     & 0.001 & 0.026
		 	\\ 		\hline  
\multirow{3}{*}{$\tau = 0.5, T=2000 $} & $\widehat{D}_T^{arb}$ ($n=10^{4}$,    &    0.007 & 0.041 & 0.076 & 0.157
	 		\\
	 		&  2 times updating $\bm{V_d}$)    &  	&	&	 	&
	 \\			
& $ \widehat{D}_T^{ker} $  & 	  0     & 0     & 0     & 0.006
		 	\\  		 	
		 	\hline
\end{tabular}
}
\end{table}

\item
Simulate 1000 samples of the DGP $\left\{ y_t \right\}$ specified as the model~\eqref{eq:CAViaR_RES_abs} with the underlying parameters are given as $\bm{\beta}_{u_t}^{R1} = [F^{-1}_{N(0,1)}(u_t), 0.2, 0.3]'$, where $\left\{ u_t \right\}\overset{i.i.d.}{\sim} \mathcal{U}(0,1)$ and $F^{-1}_{N(0,1)}(\cdot)$ is the inverse standard normal probability distribution function. Conditional $5\%$-th, $30\%$-th, $50\%$-th quantiles are estimated for each of the total $1000$ simulated samples of sample size $T$ by regressing the sample onto the full model~\eqref{eq:CAViaR_Full_AS_Wald}. 
The results of the Wald test using the adaptive random bandwidth method and the kernel method~\eqref{eq:parzon_window} are listed in Table~\ref{tab:WT_ResABs_tau_T} in which each estimated size is obtained by the percentage rejection rate among the 1000 samples of sample size $T$.
\begin{table}[h]
\centering
\caption{The size performances of the Wald test on the restricted model~\eqref{eq:CAViaR_RES_abs} to~\eqref{eq:CAViaR_Full_AS_Wald} \;  ($\bm{\beta}_{u_t}^{R1} = [F^{-1}_{N(0,1)}(u_t), 0.2, 0.3]'$, $R= [0,0,1,-1]$)}
\label{tab:WT_ResABs_tau_T}
\resizebox{\columnwidth}{!}{%
\begin{tabular}
[c]{l|l|ccccc}\hline\hline
quantile index $\tau$ \& sample size T & methods & size: $\alpha = 0.01$ & $  \alpha = 0.05$ & $\alpha = 0.10$ & $\alpha = 0.20$ \\ \hline
\multirow{3}{*}{$\tau = 0.05, T=4000 $} & $\widehat{D}_T^{arb}$ ($n=10^{4}$,    &    0.018 & 0.06  & 0.101 & 0.187		
	 		\\
	 		&  2 times updating $\bm{V_d}$)    &  	&	&	 	&
	 \\			
& $ \widehat{D}_T^{ker} $  & 0.017 & 0.064 & 0.131 & 0.235	
		 	\\ \hline
\multirow{3}{*}{$\tau = 0.05, T=2000 $} & $\widehat{D}_T^{arb}$ ($n=10^{4}$,    &   0.019 & 0.054 & 0.107 & 0.19 
	 		\\
	 		&  2 times updating $\bm{V_d}$)    &  	&	&	 	&
	 \\			
& $ \widehat{D}_T^{ker} $  &    0.035 & 0.085 & 0.14  & 0.245
			\\ \hline
\multirow{3}{*}{$\tau = 0.3, T=4000 $} & $\widehat{D}_T^{arb}$ ($n=10^{4}$,    &    0.01  & 0.058 & 0.103 & 0.187 
	 		\\
	 		&  2 times updating $\bm{V_d}$)    &  	&	&	 	&
	 \\			
& $ \widehat{D}_T^{ker} $  & 0.013 & 0.053 & 0.103 & 0.202	
		 	\\ 
		 	\hline  
\multirow{3}{*}{$\tau = 0.3, T=2000 $} & $\widehat{D}_T^{arb}$ ($n=10^{4}$,    &  0.021 & 0.061 & 0.11  & 0.184	
	 		\\
	 		&  2 times updating $\bm{V_d}$)    &  	&	&	 	&
	 \\			
& $ \widehat{D}_T^{ker} $  &  0.014 & 0.058 & 0.111 & 0.2
		 	\\ 	\hline  
\multirow{3}{*}{$\tau = 0.5, T=4000 $} & $\widehat{D}_T^{arb}$ ($n=10^{4}$,    &   0.014 & 0.062 & 0.126 & 0.221	
	 		\\
	 		&  2 times updating $\bm{V_d}$)    &  	&	&	 	&
	 \\		
& $ \widehat{D}_T^{ker} $  &  0.017 & 0.069 & 0.129 & 0.223
		 	\\ 		\hline  
\multirow{3}{*}{$\tau = 0.5, T=2000 $} & $\widehat{D}_T^{arb}$ ($n=10^{4}$,    &   	0.025 & 0.064 & 0.1   & 0.194
	 		\\
	 		&  2 times updating $\bm{V_d}$)    &  	&	&	 	&
	 \\			
& $ \widehat{D}_T^{ker} $  &   0.02  & 0.065 & 0.118 & 0.206
		 	\\  		 	
		 	\hline
\end{tabular}
}
\end{table}

\item
Simulate 1000 samples of the DGP $\left\{ y_t \right\}$ specified as the model~\eqref{eq:timevarying_ht_CAViaR_DGP_b0Bposy_restricted} with the underlying parameters are given as $\bm{\beta}_{u_t}^{R3} = [F^{-1}_{N(0,1)}(u_t), 0.2, 0.3]'$, where $\left\{ u_t \right\}\overset{i.i.d.}{\sim} \mathcal{U}(0,1)$ and $F^{-1}_{N(0,1)}(\cdot)$ is the inverse standard normal probability distribution function. Conditional $5\%$-th, $30\%$-th, $50\%$-th quantiles are estimated for each of the total $1000$ simulated samples of sample size $T$ by regressing the sample onto the full model~\eqref{eq:CAViaR_Full_AS_Wald}. 
The results of the Wald test using the adaptive random bandwidth method and the kernel method~\eqref{eq:parzon_window} are listed in Table~\ref{tab:WT_b0Nposy_varying_tau_T} in which each estimated size is obtained by the percentage rejection rate among the 1000 samples of sample size $T$.
\begin{table}[htbp]
\centering
\caption{The size performances of the Wald test on the restricted model~\eqref{eq:timevarying_ht_CAViaR_DGP_b0Bposy_restricted} to~\eqref{eq:CAViaR_Full_AS_Wald} \;  ($\bm{\beta}_{u_t}^{R3} = [F^{-1}_{N(0,1)}(u_t), 0.2, 0.3]'$, $R= [0,0,1,-1]$)}
\label{tab:WT_b0Nposy_varying_tau_T}
\resizebox{\columnwidth}{!}{%
\begin{tabular}
[c]{l|l|ccccc}\hline\hline
quantile index $\tau$ \& sample size T & methods & size: $\alpha = 0.01$ & $  \alpha = 0.05$ & $\alpha = 0.10$ & $\alpha = 0.20$ \\ \hline
\multirow{3}{*}{$\tau = 0.05, T=5000 $} & $\widehat{D}_T^{arb}$ ($n=10^{4}$,    & 0.032 &	0.069	& 0.096	& 0.17	
	 		\\
	 		&  2 times updating $\bm{V_d}$)    &  	&	&	 	&
	 \\			
& $ \widehat{D}_T^{ker} $  & 0.082	& 0.137	& 0.199	& 0.287
		 	\\ \hline
\multirow{3}{*}{$\tau = 0.05, T=2000 $} & $\widehat{D}_T^{arb}$ ($n=10^{4}$,    &       0.052 & 0.093 & 0.127 & 0.19		
	 		\\
	 		&  2 times updating $\bm{V_d}$)    &  	&	&	 	&
	 \\			
& $ \widehat{D}_T^{ker} $  & 0.143 & 0.221 & 0.271 & 0.341	
			\\ \hline
\multirow{3}{*}{$\tau = 0.3, T=5000 $} & $\widehat{D}_T^{arb}$ ($n=10^{4}$,    &  0.032	& 0.071	& 0.121	& 0.207	
	 		\\
	 		&  2 times updating $\bm{V_d}$)    &  	&	&	 	&
	 \\			
& $ \widehat{D}_T^{ker} $  & 0.073	& 0.137	& 0.207	& 0.3
		 	\\ 
		 	\hline  
\multirow{3}{*}{$\tau = 0.3, T=2000 $} & $\widehat{D}_T^{arb}$ ($n=10^{4}$,    &    0.031 & 0.063 & 0.123 & 0.204		
	 		\\
	 		&  2 times updating $\bm{V_d}$)    &  	&	&	 	&
	 \\			
& $ \widehat{D}_T^{ker} $  & 0.092 & 0.156 & 0.216 & 0.308
		 	\\ 	\hline  
\multirow{3}{*}{$\tau = 0.5, T=5000 $} & $\widehat{D}_T^{arb}$ ($n=10^{4}$,    &   0.021	& 0.055	& 0.095	& 0.188		
	 		\\
	 		&  2 times updating $\bm{V_d}$)    &  	&	&	 	&
	 \\		
& $ \widehat{D}_T^{ker} $  & 0.067	& 0.118	& 0.16	& 0.256
		 	\\ 		\hline  
\multirow{3}{*}{$\tau = 0.5, T=2000 $} & $\widehat{D}_T^{arb}$ ($n=10^{4}$,    &    0.034 & 0.069 & 0.118 & 0.208	
	 		\\
	 		&  2 times updating $\bm{V_d}$)    &  	&	&	 	&
	 \\			
& $ \widehat{D}_T^{ker} $  & 0.088 & 0.158 & 0.212 & 0.311
		 	\\  		 	
		 	\hline
\end{tabular}
}
\end{table}

\end{itemize}
\end{appendices}

%

\end{document}